\documentclass[aps,amsmath,amssymb]{revtex4}

\usepackage{graphicx}
\usepackage{bm}
\usepackage{dcolumn}
\usepackage{color}
\usepackage{pst-plot}

\newcommand{\beq}{\begin{eqnarray}}
\newcommand{\eeq}{\end{eqnarray}}

\begin{document}

\title {Nonlocal and generalized uncertainty principle black holes}
\author{Piero Nicolini}
\affiliation{Frankfurt Institute for Advanced Studies (FIAS) 
and Institut f\"ur Theoretische Physik,  Johann Wolfgang 
Goethe-Universit\"at, D-60438
Frankfurt am Main, Germany}

\date{\today}

\begin{abstract}
\noindent
In this paper we study the issue of the role of nonlocality as a possible ingredient to solve long standing problems in the physics of black holes.
To achieve this goal we analytically derive new  black hole metrics improved by corrections from nonlocal gravity actions with an entire function of the order $1/2$ and lower than $1/2$, the latter corresponding to generalized uncertainty principle corrections. This lets us extend our previous findings about noncommutative geometry inspired black holes recently recognized as nonlocal black holes due to an entire function of order higher than $1/2$. As a result we show that irrespective of the order of the function, nonlocality leads to the following properties for black hole spacetimes: i) horizon extremization also in the neutral, non rotating case; ii) black hole phase transition from a Schwarzschild phase to a positive heat capacity cooling down phase; iii) zero temperature remnant formation at the end of the evaporation process; iv) negligible quantum back reaction due to the presence of an upper bound for the Hawking temperature.
Finally we show that, in agreement with the general theory of cut off functions, a regular deSitter core accounting for the energy density of virtual gravitons replaces the curvature singularity only in the case of entire functions of order $1/2$ or higher.

\end{abstract}

\maketitle

\section{Introduction}
Nonlocality is an old age theme in the quest of regular (ultraviolet complete) theories of fundamental interactions \cite{Hei38}, but for long time it has largely been ignored  due to the initial success of regularization schemes. However, when gravity is concerned, the usual renormalization procedure runs into insurmountable difficulties and nonlocality might turn out to be an unavoidable requirement.  This is due to the fact that we expect the spacetime to drastically change its nature when probed at distances comparable to the Planck length. In such a view the conventional description in terms of smooth differential manifold would be just the low energy limit for a highly exited, granular spacetime plagued by local loss of resolution. This feature is translated as the emergence of a natural cut off, \textit{i.e.}, a fundamental length supposed to cure the usual bad short distance behavior of gravity. Therefore it is not surprising that all the major formulations of quantum gravity such as string theory, loop quantum gravity, asymptotically safe gravity, noncommutative geometry, generalized uncertainty principle encode the idea of a fundamental length and turn out to be not local in the conventional sense.
Given this background we could proceed by analyzing the nonlocal character emerging in one of the current candidate theories of quantum gravity and possibly derive phenomenological consequences from it. Even if fascinating, this route might present additional difficulties which may vary according to the limitations of each specific formulation. Alternatively one would like to study nonlocality, as a specific character by itself, irrespective of the above formulations of quantum gravity. In such a way one could hope to disclose, in a simpler way, deep conceptual features about the issue of the ultraviolet completion of gravity and, in principle, confront the conclusions with those coming from the above candidate theories of quantum gravity. This research program is so intriguing that many authors have employed nonlocal theories in a variety of contexts (for an incomplete list of examples see \cite{Pol81}). The standard procedure to implement nonlocal effects consists in considering an infinite order of derivative terms in field actions. Among the several possibilities, in this paper we want to study the effect of nonlocality on the gravitational field by assuming the form of the action recently proposed in  \cite{GHS10,Mof10,MMN11}, \textit{i.e.},
 \begin{eqnarray}
S_{\mathrm{grav}} 
=\frac{1}{ 16\pi G} \int \, d^{4} x \, \sqrt{-g} \ {\cal R}(x),
\label{uvfeh}
\end{eqnarray}
where in place of the Ricci scalar $R(x)$ we have 
\begin{equation}
{\cal R}(x)=\int d^4y \ \sqrt{-g} \ {\cal A}^2(x-y) R(y)
\end{equation}
with  ${\cal A}^2(x-y)$ a  bi-local distribution defined by
\begin{equation}
{\cal A}^2(x-y)={\cal A}^2\left(\Box_x \right)\ \delta^{(4)}(x-y).
\end{equation}
Here ${\cal A}\left(\Box_x \right)$ is an operator of $\Box_x=\ell^2 g_{\mu\nu}\nabla^\mu\nabla^\nu$, the dimensionless generally covariant D'Alembertian operator, and $\ell$ a free parameter which fixes the length scale of the theory. The above action can be seen as the leading term of what recently proposed   in the context of a quantum theory of gravity with higher derivative corrections to the Einstein-Hilbert action \cite{Mod11}.
The general strategy of theories of this kind is to find a suitable form of ${\cal A}\left(\Box_x \right)$ so that the Euclidean momentum space function ${\cal A} (-p^2)$ plays the role of the cutoff function in the quantum gravity perturbation theory expanded against a fixed background spacetime at all orders. 

To reach the above goal we need to restrict the functional class where  ${\cal A}$ can be chosen.
 For notational convenience, let us start by writing the Euclidean momentum space representation of ${\cal A}$ as
${\cal A}(z)$
where $z=-\ell^2 p^2$.
According to \cite{Efi67}, the first requirement for ${\cal A}(z)$ is to be an entire function when treated as  a function of the complex variable $z$. Otherwise any singularities of the function ${\cal A}(z)$ for finite $z$ will lead to the appearance of some additional non-physical singularities in (\ref{uvfeh}).  Then we distinguish three possible cases. First, the function ${\cal A}(z)$ is an entire function of order $\alpha<1/2$, {\it i.e.}, 
\begin{equation}
|{\cal A}(z)|<\mathrm{e}^{a|z|^\alpha}
\end{equation}
where $a$ is a positive number. For these functions there is not any direction in the complex plane $z$, along which they could decrease. Consequently, they cannot play the role of the cutoff functions and in this case it is impossible to make the perturbation theory finite. A second case occurs when the function ${\cal A}(z)$ is an entire function of order $1/2$, {\it i.e.}, 
\begin{equation}
|{\cal A}(z)|\leq \mathrm{e}^{a|z|^{1/2}}.
\end{equation}
Such a function may have only one direction in the complex plane along which it may decrease and provide a consistent cutoff. Finally the third choice is an entire function of order higher than $1/2$, {\it i.e.},
\begin{equation}
|{\cal A}(z)|\leq \mathrm{e}^{h(|z|)}
\end{equation}
where $h(|z|)$ is a positive function satisfying the condition $h(|z|)>a|z|^{1/2}$ at $|z|\to\infty$ for any $a> 0$. Such a function may have some regions where it converges  when $|z|\to\infty$ so that it may be chosen as a
cutoff function for the construction of a finite unitary perturbation theory. 

In this paper we draw our attention to nonlocal gravity as emerging from virtual graviton exchange. To this purpose we recall that the family of noncommutative geometry inspired black holes (NCBHs) \cite{Nic05,NSS06b,NiS10,NiT11} (for a review see \cite{Nic09}) corresponds to black hole solutions of nonlocal gravity with an operator ${\cal A}$ represented in terms of an entire function of order higher than $1/2$ \cite{MMN11}, {\it i.e.},
\begin{equation}
{\cal A}\left(p^2 \right)=\exp\left(\ell^2 p^2 /2 \right).
\label{ncentire}
\end{equation}
For the strong nonlocal character implemented by entire functions of this kind, NCBHs exhibit intriguing features. First, NCBHs consistently reproduce at scales larger than $\ell$ the behaviour of the corresponding Einstein gravity black holes. Second, NCBHs enjoy regularity properties at the origin, where in place of the conventional curvature singularity a regular deSitter core accounts for the nonlocal, fluctuating character of the manifold in terms of quantum vacuum energy associated to virtual gravitons. Local violations of energy conditions at the origin confirm the departure from a classical description of the gravitational field at short scales. Third, NCBHs have an improved thermodynamics: the Hawking temperature, even for the static, neutral case, admits a maximum followed by a positive heat capacity, stable, terminal phase of the evaporation \cite{BMS08}. In addition NCBHs have companion geometries like regular collapsing shells \cite{OhP10} and traversable wormholes \cite{Gal09}.

Despite their attractive features, NCBHs might not  be the only kind of nonlocal black holes. As a consequence we want to study the case of entire functions different from (\ref{ncentire}).
The goal of this investigation is twofold: on the one hand we would like to know if the new features of NCBHs depend on the specific choice of the above entire function or can be considered as general characters of nonlocality; on the other hand we would like to release the strong requirement of entire function of order higher than $1/2$ to study milder and more economic nonlocal modifications of gravity. Not surprisingly, in doing so we will show that not only noncommutative geometry but also a theory based on the generalized uncertainty principle (GUP) is nothing else than a particular realization of a nonlocal theory. 

The paper is organized as follows: in section \ref{uno} we derive the general form of nonlocal corrections to black hole metrics, in section \ref{due} we analyze the case of an entire function of order $1/2$, in section \ref{tre} we study a model of GUP black hole and finally in section \ref{quattro} we draw the conclusions by comparing all the existing families of black holes derived up to now in this way.

\section{Nonlocal modifications to the metric tensor}
\label{uno}
To derive nonlocal effects on the gravitational interaction we need an action to describe how the nonlocal geometry, modeled by (\ref{uvfeh}), behaves when coupled to a static matter source, {\it i.e.}, 
\begin{equation}
S_{\mathrm{tot}}=S_{\mathrm{grav}}+S_{\mathrm{matt}}
\label{totaction}
\end{equation}
where the matter action $S_{\mathrm{matt}}$ will be specified below. The field equations can be obtained by varying (\ref{totaction}) with respect to the metric $g_{\mu\nu}$. By neglecting surface terms coming from the variation of the generally-covariant D'Alembertian, we find the nonlocal version of Einstein equations \cite{GHS10,Mof10,MMN11}
\begin{eqnarray}
{\cal A}^{2}\left(\Box \right)\left(R_{\mu \nu} - \frac{1}{2} g_{\mu \nu}R \right)= 8 \pi G
 T_{\mu \nu}.
\label{e0}
\end{eqnarray}
We notice that these equations can be cast in an equivalent form. The advantage of the Euclidean momentum space representation is that ${\cal A}$ becomes an elliptic operator and therefore has a unique, well defined inverse. One can conveniently work in Euclidean momentum space and rotate back to pseudo Euclidean spacetime at the end of the calculation. As a result one finds
\begin{eqnarray}
R_{\mu \nu} - \frac{1}{2} g_{\mu \nu}R = 8 \pi
G\ {\mathfrak T}_{\mu\nu},
\label{e1}
\end{eqnarray}
where the divergence free tensor
\begin{equation}
{\mathfrak T}_{\mu\nu}\equiv{\cal A}^{-2}\left(\Box\right) T_{\mu \nu}.
\label{exoticstress}
\end{equation}
Eq. (\ref{e1}) has the familiar form of Einstein gravity coupled to a generalized source term ${\mathfrak  T}_{\mu\nu}$, while (\ref{e0}) describes nonlocal gravity equations coupled to ordinary matter $T_{\mu\nu}$. The two forms are equivalent. As a static source we consider the case of a pressure-less static fluid at the origin whose action is given by
\begin{equation}
S_{\mathrm{matt}}=-\int \, d^{4} x \, \sqrt{-g} \ \rho_0(x),\ \quad \rho_0(x)=\frac{M}{\sqrt{-g}} \int \, d\tau  \ \delta\left(x-x(\tau)\right)
\end{equation}
where $\rho_0(x)$ is the energy density describing a massive, point-like particle. The corresponding energy momentum tensor reads
\begin{eqnarray}
 T^0\,_0=-\frac{M}{4\pi\, r^2}\delta ( r ),\label{t00}
\end{eqnarray}
and the spacetime, solution of either (\ref{e0}) and (\ref{e1}), enjoys the usual static, spherically symmetric form
\begin{equation}
ds^2=-f(r)~dt^2 + f^{-1}(r)~dr^2+r^2d\Omega^2,
\end{equation}
where the coefficient $f(r)$ has the form
\begin{equation}
f(r)=1-\frac{2{\cal G}(r) M }{r}.
\label{ef}
\end{equation}
The function ${\cal G}(r)$ is the unknown and incorporates all the sought effects of nonlocality. In the limit  $r\gg\ell$, the function ${\cal G}(r)$ is subject to the requirement of matching Newton's constant, ${\cal G}(r)\to G$. Before solving the above equations one has to prescribe the  form of ${\cal A}\left(\Box_x \right)$.

\section{Static, spherically symmetric black hole solution for $\alpha=1/2$}
\label{due}
%

We start by considering the entire functions of order $\alpha= 1/2$, {\it i.e.},
\begin{equation}
{\cal A}(p^2)=\exp(\ell p/2).
\label{onehalf}
\end{equation}
To determine ${\cal G}(r)$, it is more convenient to follow the gravity field equations in the form (\ref{e1}) and derive  ${\mathfrak T}_{\mu\nu}$ by applying ${\cal A}^{-2}$ on $T_{\mu\nu}$. 
For later convenience we temporarily adopt free falling Cartesian-like coordinates and we calculate
\begin{equation}
{\mathfrak T}^0\,_0= - M {\cal A}^{-2}\left(\Box\right)\delta(\vec{x})\equiv - M \mathfrak{g}(\vec{x}).
\end{equation}
As a first step we introduce the following Schwinger representation 
\begin{equation}
\ln {\cal A}(\Box)=\frac{1}{2}\left(-\Box\right)^{\frac{1}{2}}=\frac{1}{2\Gamma(-\frac{1}{2})}\int_0^\infty ds \ s^{-\frac{3}{2}}{\mathrm e}^{-s(-\Box)}.
\end{equation}
Second we apply  the above operator on the Dirac delta, {\it i.e.},
\begin{eqnarray}
\ln {\cal A}(\Box)\ \delta(\vec{x})&=&\frac{1}{2\Gamma(-\frac{1}{2})}\int_0^\infty ds \ s^{-\frac{3}{2}}{\mathrm e}^{-s(-\nabla^2)} \frac{1}{(2\pi)^3}  \int d^3p \ \mathrm{e}^{i\vec{x}\cdot\vec{p}} \nonumber \\
&=&\frac{1}{(2\pi)^3}  \int d^3p \ \left(\frac{1}{2}\ \ell|\vec{p}|\right)\mathrm{e}^{i\vec{x}\cdot\vec{p}}. \nonumber
\end{eqnarray}
As a result one obtains the action of the operator ${\cal A}^{-2}\left(\Box\right)$, {\it i.e.},
\begin{eqnarray}
{\cal A}^{-2}\left(\Box\right)\delta(\vec{x}) = 
\frac{1}{(2\pi)^3} \int d^3p \ \mathrm{e}^{-\ell |\vec{p}|} \ \mathrm{e}^{i\vec{x}\cdot\vec{p}}.
\label{aaction} 
\end{eqnarray}
By integrating the r.h.s. of (\ref{aaction}) one gets
\begin{equation}
\mathfrak{g}(\vec{x})=\frac{1}{\pi^2}\frac{\ell}{(\vec{x}^2+\ell^2)^2}.
\label{enprof}
\end{equation}
The function $\mathfrak{g}(\vec{x})$ is a Student's t-distribution with $3$ degrees of freedom \cite{AbS65}. It admits just $2$ moments. 
We recall that physically the quantity $\rho(\vec{x})\equiv M \mathfrak{g}(\vec{x})$ is the energy density, which is smeared out for the effect of nonlocality. Therefore the mass $M$ is no longer concentrated into a point but it is ``diluted'' according to 
\begin{equation}
{\cal M}(r) = - 4\pi \int dr r^2\ {\mathfrak T}^0\,_0 .
\label{mass}
\end{equation}
This mass cumulative distribution ${\cal M}(r)$ lets us solve Einstein equations in terms of 
the metric component written as $f(r)=1-2{\cal M}(r)G/r$. By equating the latter with (\ref{ef}),
one can determine the initial unknown 
\begin{equation}
{\cal G}(r)= 4\pi G \int dr r^2\ \mathfrak{g}(r).
\end{equation}
The final step is to calculate the above integral. One finds
\begin{equation}
{\cal G}(r)=G\, \left(\frac{2}{\pi} \arctan(r/\ell) - \frac{2}{\pi}\frac{r/\ell}{(1+(r/\ell)^2)}\right).
\label{mass2}
\end{equation}
  \begin{figure}
 \begin{center}
\includegraphics[height=7.0cm]{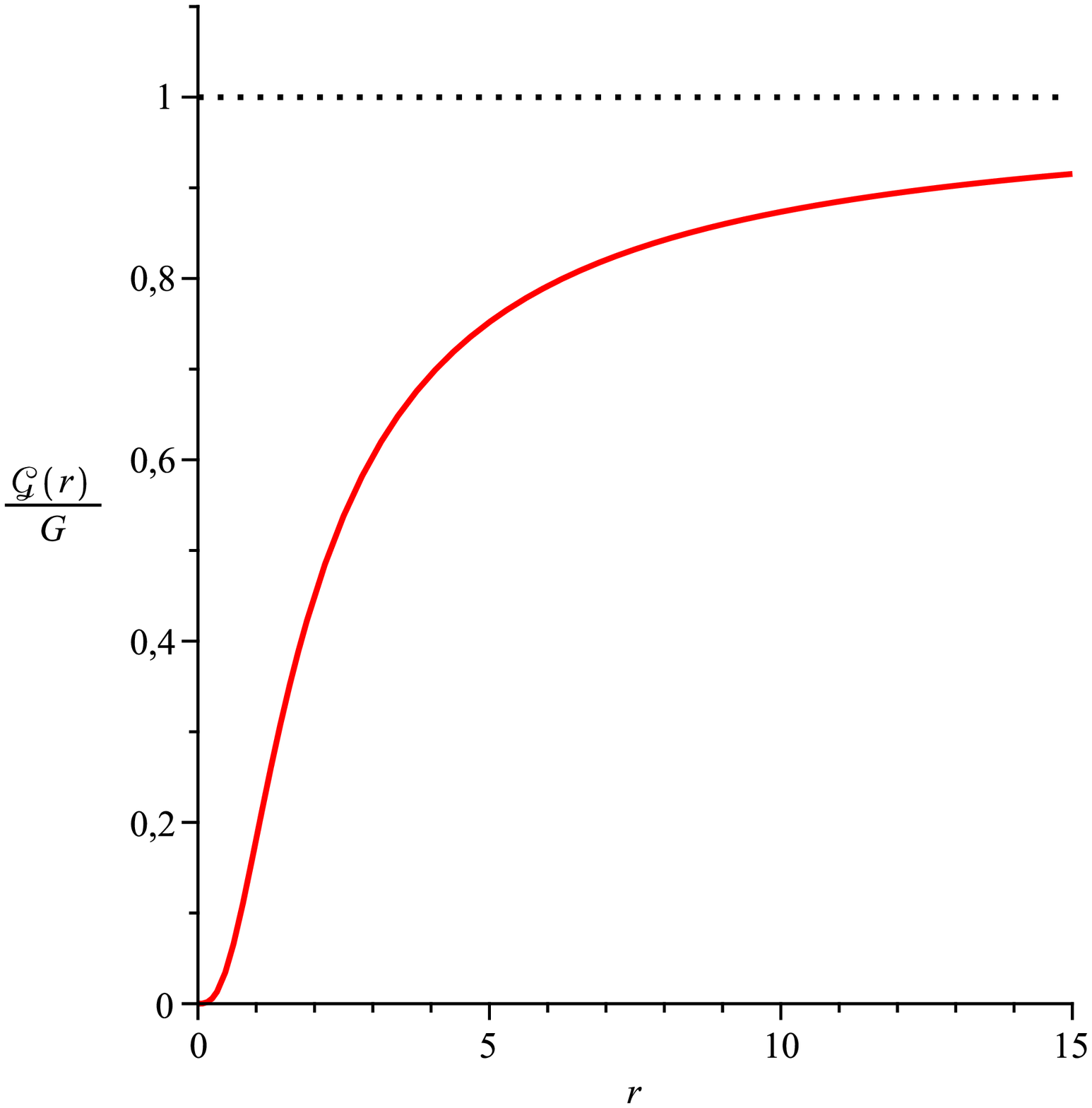}
  \includegraphics[height=7.0cm]{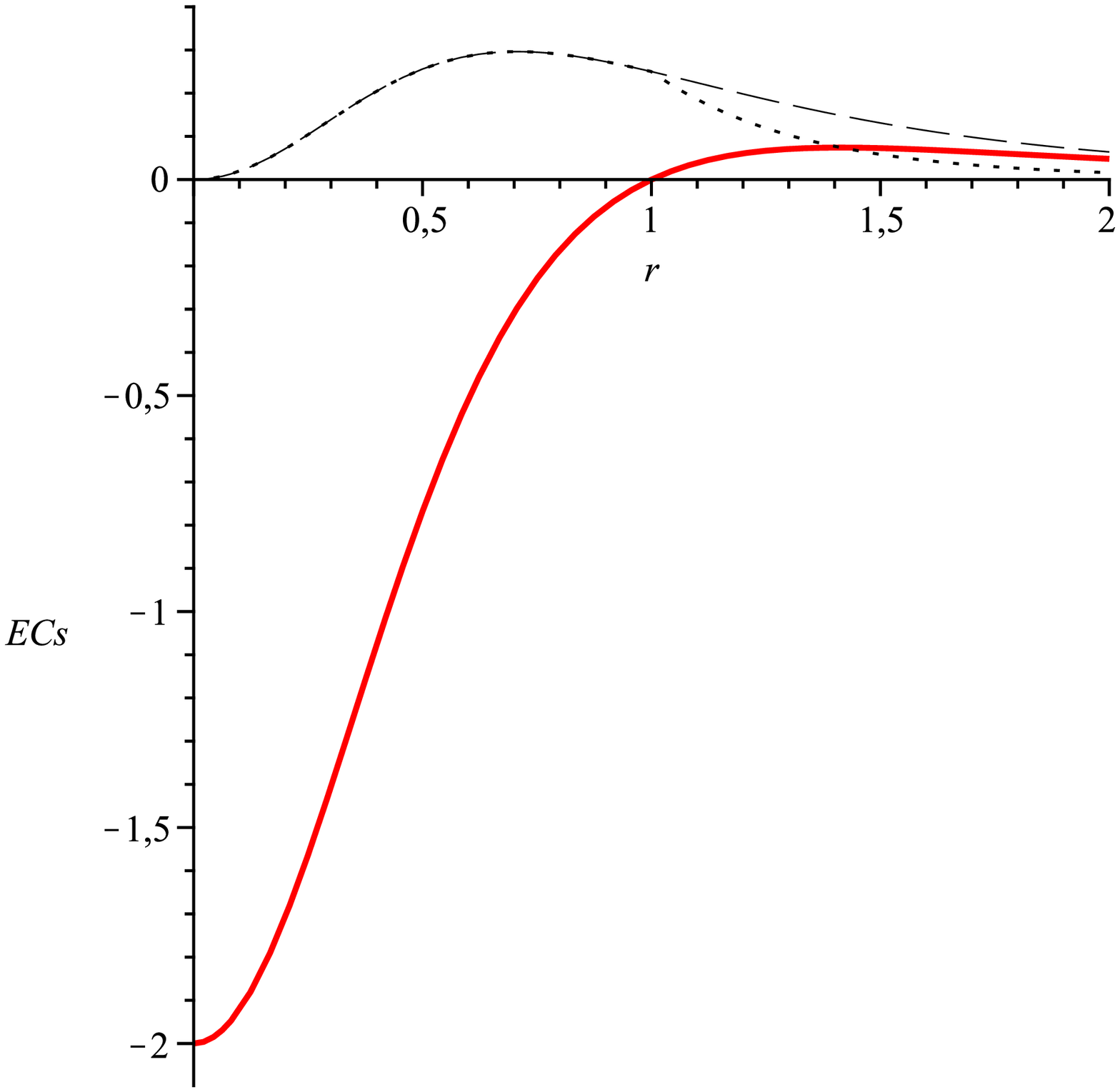}
  \hspace{0.2cm}
      \caption{\label{figG} On the left we have the function ${\cal G}(r)/G$ (solid line), while on the right the solid curve is the function $(\rho +p_{\mathrm{rad}}+2p_\perp)$ (strong energy condition), the dotted curve is the fucntion $(\rho-|p_\perp|)$ (dominant energy condition) and the dashed curve is $(\rho +p_\perp)$ (null energy condition). In both plots we set $\ell=1$.}
     \end{center}
  \end{figure}
In Fig. \ref{figG}, we have the plot of ${\cal G}(r)/G$. We see that nonlocal gravity matches Einstein gravity in the infrared limit and it is asymptotically free.
This can be checked by expanding (\ref{mass2}) for $r\gg \ell$ 
\begin{equation}
{\cal G}(r)\approx G \ \left[ 1- \frac{4}{\pi}\left(\frac{\ell}{r}\right)+\frac{8}{3\pi}\left(\frac{\ell}{r}\right)^3\right],
\end{equation}
which confirms that at large distances the solution (\ref{ef}) tends to the Schwarzschild geometry.
The above expression can be used to predict experimentally testable deviations to  Newton's law, which turns out to be of the form
\begin{equation}
\phi(r)=-G\frac{M}{r}\left( 1+ \beta_k\left\{\frac{1\ \mathrm{mm}}{r}\right\}^{k-1}\right).
\label{NL}
\end{equation}
Current experimental limits on $\beta_k$ would not give tighter constraints than $\ell\lesssim 1$ $\mu$m \cite{Hoy04}.  On the other hand we can expand (\ref{mass2}) for $r\ll \ell$ to get
\begin{equation}
{\cal G}(r)\approx \frac{4}{3\pi}\ \left(\frac{r}{\ell}\right)^3 \ G
\end{equation}
which confirms the asymptotic freedom of the coupling. By substituting the above expression into (\ref{ef}), we get
\begin{eqnarray}
\hspace{-0.15cm} ds^2\approx - \left(1-\frac{1}{3}\Lambda_{\mathrm{eff}} r^2\right)dt^2 + \frac{dr^2}{\left(1-\frac{1}{3}\Lambda_{\mathrm{eff}} r^2\right)}+r^2\Omega^2.
\end{eqnarray}
This is a deSitter line element with an effective cosmological constant
$\Lambda_{\mathrm{eff}} = 8M    G  \ell^{-3}\pi^{-1}$.
The regularity of the solution proves that our choice of entire function (\ref{onehalf}) leads to an ultraviolet complete theory of gravity. 
On physical grounds the picture is the following. Nonlocality makes the gravitational interaction weaker and weaker as the energy scale increases. This is a net effect due to the emergence of repulsive quantum gravitational fluctuations in the UV region. 
The deSitter core at the origin accounts for the mean value of the energy of these fluctuations. This fact becomes even more evident if we study pressure terms we need to balance the generalized energy momentum tensor, {\it i.e.}, $\nabla_\mu {\mathfrak T}^{\mu\nu}=0$. From the conservation equation we get
\begin{equation}
\partial_r {\mathfrak T}_r^{\ r}=-\frac{1}{2}f^{-1}(r)\partial_r f(r)\left({\mathfrak T}_r^{\ r}-{\mathfrak T}_0^{\ 0}\right)-\frac{2}{r}\left({\mathfrak T}_r^{\ r}-{\mathfrak T}_\theta^{\ \theta}\right)
\label{TOV}
\end{equation}
where ${\mathfrak T}_\nu^{\ \mu}=\mathrm{diag}\left(-\rho(r), p_\mathrm{rad}(r), p_\perp(r), p_\perp(r)\right)$.
By solving (\ref{TOV}), we get that both the radial pressure  $p_\mathrm{rad}(r)=-M\mathfrak{g}(r)$ and the angular pressure $
p_\perp(r)=-M\mathfrak{g}(r)-\frac{r}{2}\ M\partial_r \mathfrak{g}(r)
$
turn out to be negative. As a further evidence of the non-classical character of the deSitter core it is useful to study energy conditions (see Fig. \ref{figG}). In a spherical cap around the origin the strong energy condition is violated 
\begin{equation}
\rho(r) +p_\mathrm{rad}(r)+2p_\perp(r) = \frac{2M\ell}{\pi^2}\frac{r^2-\ell^2}{\left(r^2+\ell^2\right)^3}< 0, \quad \mathrm{if}\ r<\ell .
\end{equation}
The locally exotic character of the solution is consistent with the repulsive effect of nonlocal gravity which sustains the generalized matter energy profile $\rho(r)$ and prevents a collapse into a point like profile $M\delta(r)$.

   \begin{figure}
 \begin{center}
 \includegraphics[height=7.0cm]{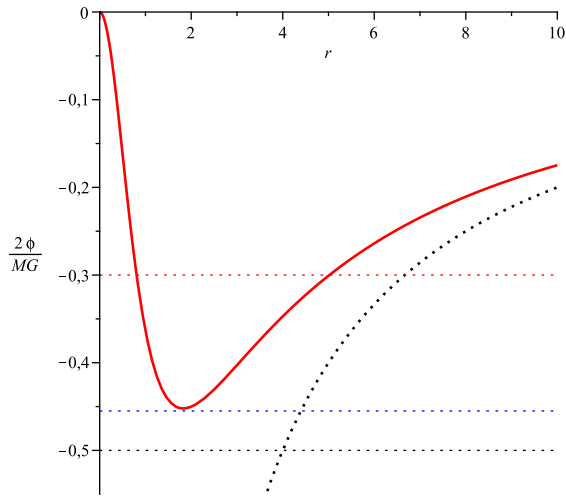}
 \hspace{0.2cm}
      \caption{\label{figphi}The function $y=2\phi (r)/MG=-2{\cal G}(r)/(Gr)$ (solid line) is plotted in comparison with the classical curve $y=-2/r$, by setting $\ell=1$. Intersections with horizontal curves $y=-1/MG$ give horizon radii. 
 }
 \end{center}
  \end{figure}
To study the horizon equation one may wish to solve the equation $f(r_H)=0$. However, for fixed $\ell$, this is a parametric equation depending on the mass parameter $M$ and cannot in general be solved in a closed form. Alternatively, in Fig. \ref{figphi}, we have the function $2\phi (r)/MG$ (solid line), where $\phi(r)=-{\cal G}(r)M/r$ is the new profile of the Newtonian potential. Horizontal lines are curves with constant ordinate $-1/MG$. We see that in the classical case (dotted line), the Newtonian potential has a hyperbolic behaviour and always admit one intersection with horizontal lines for any value of $M$. This corresponds to the event horizon of the Schwarzschild solution at $r_g=2MG$. Conversely in the nonlocal gravity case the Newtonian potential reaches an extremum (a minimum) at $r_0$ before vanishing at the origin. This means that intersections can occur only if horizontal lines lie above the minimum of the solid line in Fig. \ref{figphi}. More specifically there exists a value $M_0$ for the mass parameter such that
\begin{itemize}
\item for $M<M_0$ there is no intersection, corresponding to the case of a regular manifold without horizons;
\item for $M>M_0$ there are two intersections, corresponding to an inner horizon $r_-$ and an outer horizon $r_+$:
\item for $M=M_0$ there is just one intersection, corresponding to a single degenerate horizon $r_0$.
\end{itemize}
\begin{figure}[ht]
\begin{center}
\includegraphics[height=10cm,angle=0]{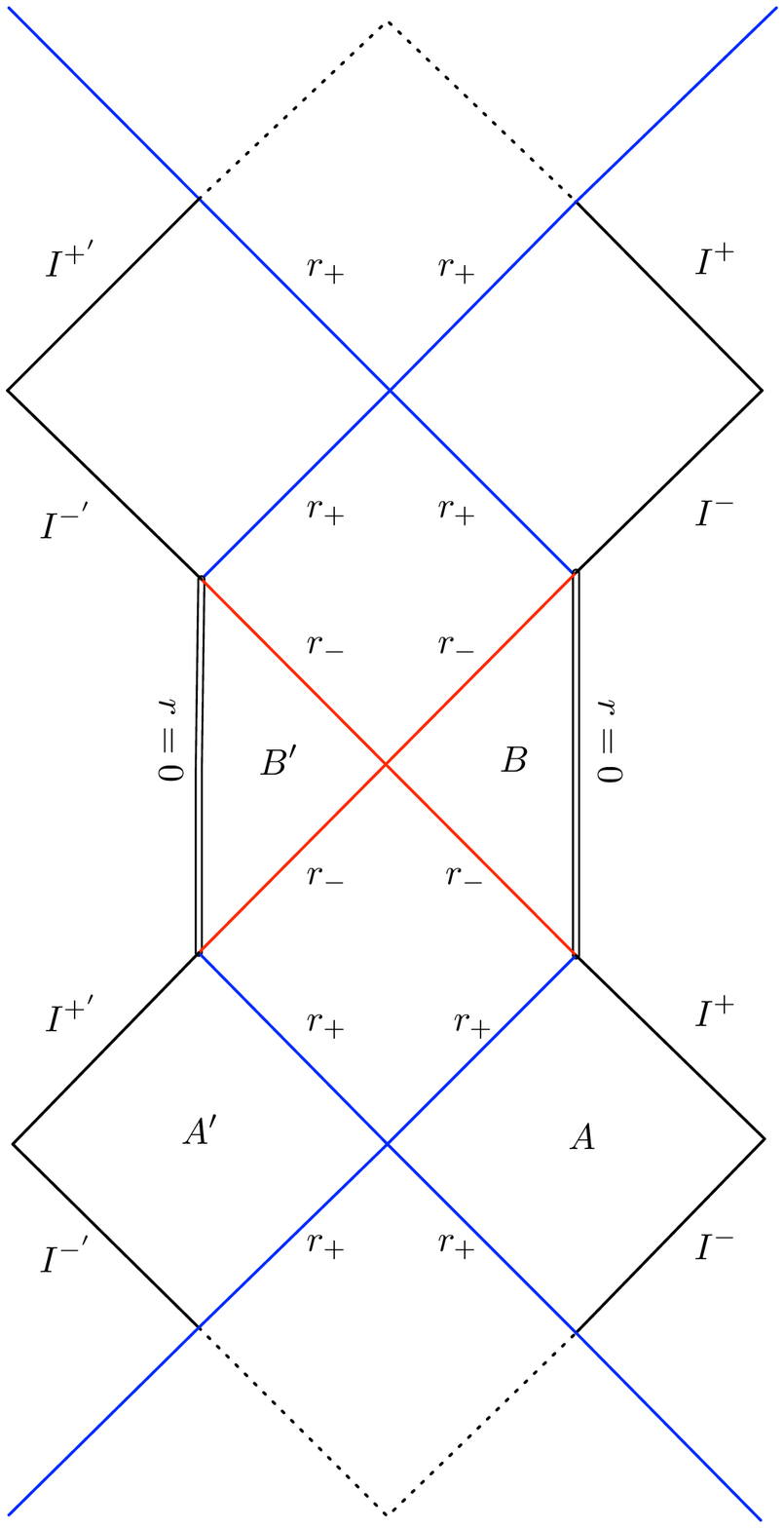}
\caption{\label{penrose}The solution admits one, two or no horizons depending on $M$. In the case of two horizons,  $f(r_\pm)=0$, the Penrose diagram resembles the Reissner-Norstr\"{o}m geometry, except for the origin where a regular deSitter core lies in place of the curvature singularity.}
\end{center}
\end{figure}
By a numerical estimate we obtain $M_0\approx 2.21\  \ell/G$ and $r_0\approx 1.83\ \ell$.  
We notice that for large masses horizontal lines in Fig. \ref{figphi} descend and outer horizons approach the gravitational radius $r_+\approx r_g$.
The global structure of the solution looks like that of a Reissner-Nordstr\"{o}m black hole with a regular origin in place of the curvature singularity (see Fig. \ref{penrose}). The regularity of the manifold discloses further properties of the solution. If one consider negative values of the radial coordinate, one does not get a mere analytical continuation of spacetime with positive $r$. Since at the origin, the spacetime is locally flat, the  solution with positive $r$ is geodesically complete and therefore negative $r$ values describe an additional distinct spacetime. On the other hand, being $\arctan(r/\ell)$ an odd function, the  solution with negative $r$ is the mirror symmetric of  the spacetime with  positive $r$.
   \begin{figure}
 \begin{center}
 \includegraphics[height=7.0cm]{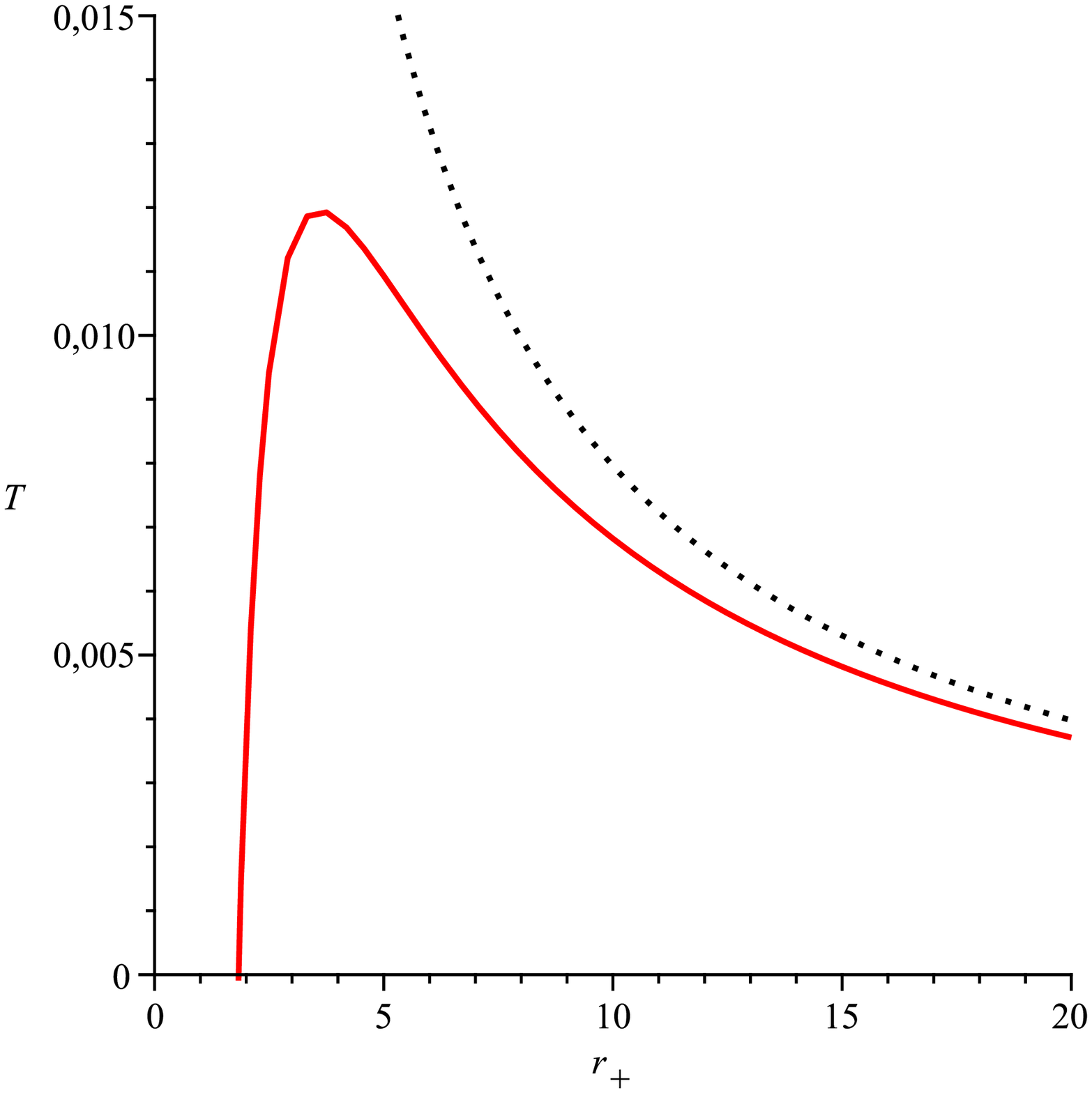}
 \includegraphics[height=7.0cm]{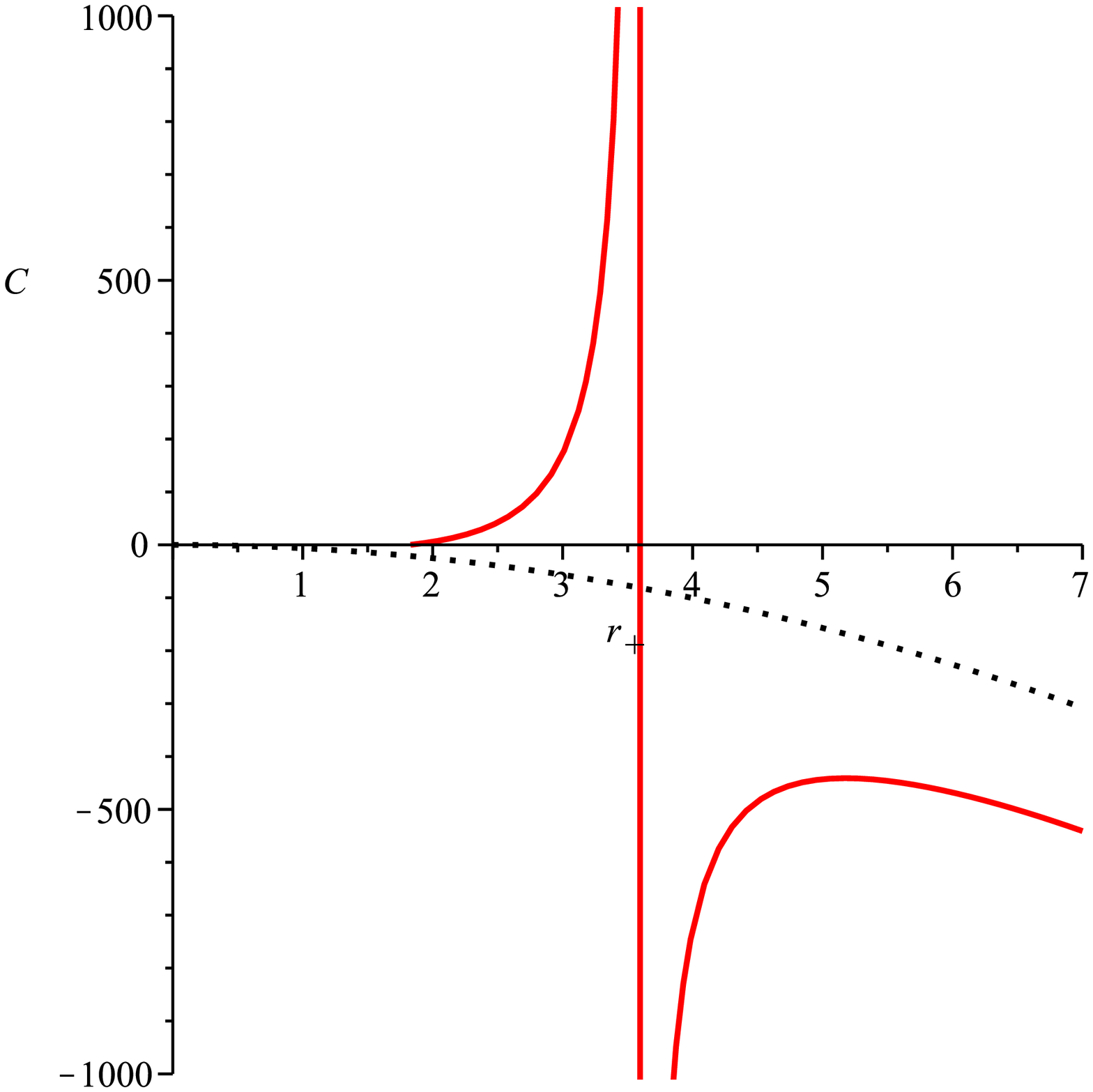}
 \hspace{0.2cm}
      \caption{\label{figtemp} On the left we have the black hole temperature and on the right its heat capacity. In both plots solid lines represent the new behaviors while dotted lines are the classical results. In addition we set $\ell=1$.
 }
 \end{center}
  \end{figure}
  
To analyze the thermodynamic properties of the black hole, we perform a Wick rotation
\begin{equation}
ds^2 = -f(r)~dt^2 + f^{-1}(r)~dr^2+r^2d\Omega^2 ~~\longrightarrow ~~ ds_E^2 = f(r)~d\tau^2 + f^{-1}(r)~dr^2+r^2d\Omega^2 .
\end{equation}
The periodicity of the imaginary time yields the following profile for the black hole temperature \cite{Nic10}
\begin{equation}
T=\frac{1}{4\pi r_+}\left( 1-r_+\frac{{\cal G}^\prime(r_+)}{{\cal G}(r_+)}\right).
\label{HT}
\end{equation}
The entropy of the black hole can be calculated by integrating $dS=dM/T$ and reads 
\begin{eqnarray}
S=\frac{1}{4}\ \int_{A_0}^{A_+} \frac{dA_{H}}{{\cal G}(r_H)}
\label{BHE}
\end{eqnarray}
where $A_H=4\pi r_H^2$ is the event horizon area, and $A_0$ and $A_+$ are the same quantity calculated in $r_0$ and $r_+$ respectively. In (\ref{HT}) the term ${\cal G}^\prime/{\cal G}$ provides deviations with respect to the Hawking's result, which is reproduced for constant ${\cal G}=G$. By looking for the extremum of Netwonian potential 
$\phi^\prime = (GM/r^2)\left( 1-r{\cal G}^\prime/{\cal G}\right)$,
one obtains that the temperature vanishes at $r_0$. This is in agreement with the usual thermal behavior of extremal black holes and confirms early expectations about a possible evaporation switching off of the Schwarzschild black hole  by taking into account quantum back reaction effects \cite{BaB88}. However in our case the mechanism of horizon extremization has a different explanation: the switching off is driven by ${\cal G}^\prime/{\cal G}$ which models the  modifications of the nonlocal geometry at high energies. In such a view the Schwarzschild geometry is a ``rigid'' background which neglects high energy quantum modifications of spacetime.   
By calculating ${\cal G}^\prime$ one can display the final formula for the temperature
\begin{equation}
T=\frac{1}{4\pi r_+}\left( 1-2\frac{(r_+/\ell)^3}{(1+(r_+/\ell)^2)^2}\left(\arctan(r_+/\ell) - \frac{r_+/\ell}{(1+(r_+/\ell)^2)}\right)^{-1}\right)
\end{equation}  
whose plot is in Fig. \ref{figtemp}. For large $r_+\approx r_g$ the nonlocal corrections are suppressed and one recovers the usual $1/r_+$ behaviour. At $r_+=r_{\mathrm{max}}\approx 3.60\ \ell$, the temperature reaches a maximum $T_{\mathrm{max}}=1.19\times 10^{-2}\ \ell/G$, before vanishing at $r_+=r_0\approx 1.83 \ \ell$, an extremal black hole remnant configuration. As a consequence we obtain a ratio $T/M<T_{\mathrm{max}}/M_0\approx 5.38\times 10^{-3}$, which implies that the vacuum energy density of quantum fields propagating on the black hole spacetime cannot significantly modify the background geometry (\ref{ef}). In other words the black hole is so cold that the sole term ${\cal G}^\prime/{\cal G}$ is sufficient to consistently describe all metric modifications during the whole evaporation process.

Eq. (\ref{BHE}) is a consistent generalization of the well known area law. The problem is that we cannot determine the exact form of the entropy analytically. However we can write (\ref{BHE}) as
\begin{equation}
S=\frac{1}{4}\left(\frac{A_+}{{\cal G}(r_+)}-\frac{A_0}{{\cal G}(r_0)}\right)  +\frac{1}{4}\ \int_{r_0}^{r_+} \frac{{\cal G}^\prime(r_H)}{{\cal G}^2(r_H)} A_{H} \ dr_H
\label{BHE2}
\end{equation}
and evaluate it by an expansion in powers $(r_H/\ell) > 1$ (being $r_H\geq r_0>\ell$). As a result one finds (see (\ref{gexp}))
\begin{equation}
S\approx \frac{1}{4{\cal G}(r_+)}\ A_+ + O(\ell/r_+),
\label{entrapp}
\end{equation}
which confirms our expectations: up to subleading correction, the area law is preserved in form by a mere substitution of the classical coupling constant $G$ with the nonlocal one ${\cal G}(r_+)$. 

A further inspection in the effects of nonlocality can be made by looking at the black hole heat capacity $C\equiv dM/dT$, that can be written as
\begin{equation}
C(r_+)=T\ \left(\frac{dS}{dr_+}\right)\left(\frac{dT}{dr_+}\right)^{-1}.
\label{heatcap1}
\end{equation} 
This relation highlights the black hole exchanges of energy with the environment.
Both $T$ and $dS/dr_+ ( \sim dA_+ / {\cal G}(r_+))$ are positive defined, while $dT/dr_+$ is negative for $r_+>r_{\mathrm{max}}$, vanishing for $r_+=r_{\mathrm{max}}$ and positive for $r_+<r_{\mathrm{max}}$. As a result for large $r_+$ the heat capacity  tends to the classical negative value $C=-2\pi r_+^2/G$ since nonlocal effect are negligible at these scales. By decreasing $r_+$ the heat capacity grows negatively, departing from the usual Schwarzschild behaviour. At $r_+=r_{\mathrm{max}}$ the heat capacity has a vertical asymptote and switches sign. This means that a phase transition occurs from an unstable thermodynamic phase ($C<0$) to stable one ($C>0$). In such a stable phase, the evaporation becomes a disadvantageous process, slowing down in the vicinity of the zero temperature configuration.  From a thermodynamic viewpoint black hole remnants are asymptotic stable configurations, that cannot exchange energy and evaporate ($T=C=0$). In Fig. (\ref{figtemp}) we see that on the left of the asymptote the heat capacity descends intersecting the $r_+$-axis in $r_+=r_0$. Then the curve stops since horizons do not form at radii smaller than $r_0$.
The full expression for the heat capacity $C(r_+)$ can be found in (\ref{heatcap}).

\section{Generalized uncertainty principle black holes}
\label{tre}

Generalized uncertainty relations have been postulated on the basis of string theory arguments to account the emergence of effective ultraviolet cutoff {\it i.e.} a minimal observable length (for a recent pedagogical review see \cite{SNB12}). In one dimension the simplest generalized uncertainty relation has the form
\begin{equation}
\Delta x\Delta p\ge \frac{\hbar}{2}\left(1+\beta(\Delta p)^2+\gamma\right)
\end{equation}
where $\beta$ and $\gamma$ are positive and independent. While in ordinary quantum mechanics $\Delta x$ can be made arbitrarily small, this is no longer the case for the above equation. Due to the presence of the term $\beta(\Delta p)^2$ which grows faster than the l.h.s. term, there exist a nonzero minimal uncertainty $∆x_0$ in position. This result can be generalized to a $n$ dimensional Euclidean manifold. By postulating commutation relations like
\begin{eqnarray} 
\left[\hat{q}^i, \hat{p}_j\right]&=&i\delta^i_{j}\left(1+ \beta\hat{ \vec{p} }^2\right),\nonumber \\
\left[\hat{p}_i, \hat{p}_j\right]&=&0,
\end{eqnarray}
one obtains that coordinate operators fulfill the following noncommutative relation
\begin{eqnarray} 
\left[\hat{q}^i, \hat{q}^j\right]&=&-2i\ \beta \left(1+\beta \hat{ \vec{p} }^2\right)\hat{L}^{ij},
\label{NCGUP}
\end{eqnarray}
where $\hat{L}_{ij}$ is the generator of rotations. As a consequence the integration measure in momentum space
\begin{equation}
1=\int\frac{d^n p}{\left(1+\beta \vec{p} ^2\right)}\left|p\right>\left<p\right|
\end{equation}
becomes suppressed in the UV region \cite{KMM95}. The repercussions of GUP on the physics of black holes have been largely investigated (see for instance early contributions, {\it e.g.} \cite{APS01,Sca99}, and recent developments, {\it e.g.} \cite{CMP11}): the general conclusion is that the introduction of the UV cut off $\sqrt{\beta}$ can prevent black holes from evaporating completely with a consequent remnant formation. The latter would play a significant role in cosmology as a candidate to being the primary source of dark matter \cite{ChA03}. These considerations are based on the implementation of a generalized uncertainty relation to particles emitted by the black holes. 

 \begin{figure}
 \begin{center}
  \includegraphics[height=7.5cm,angle=0]{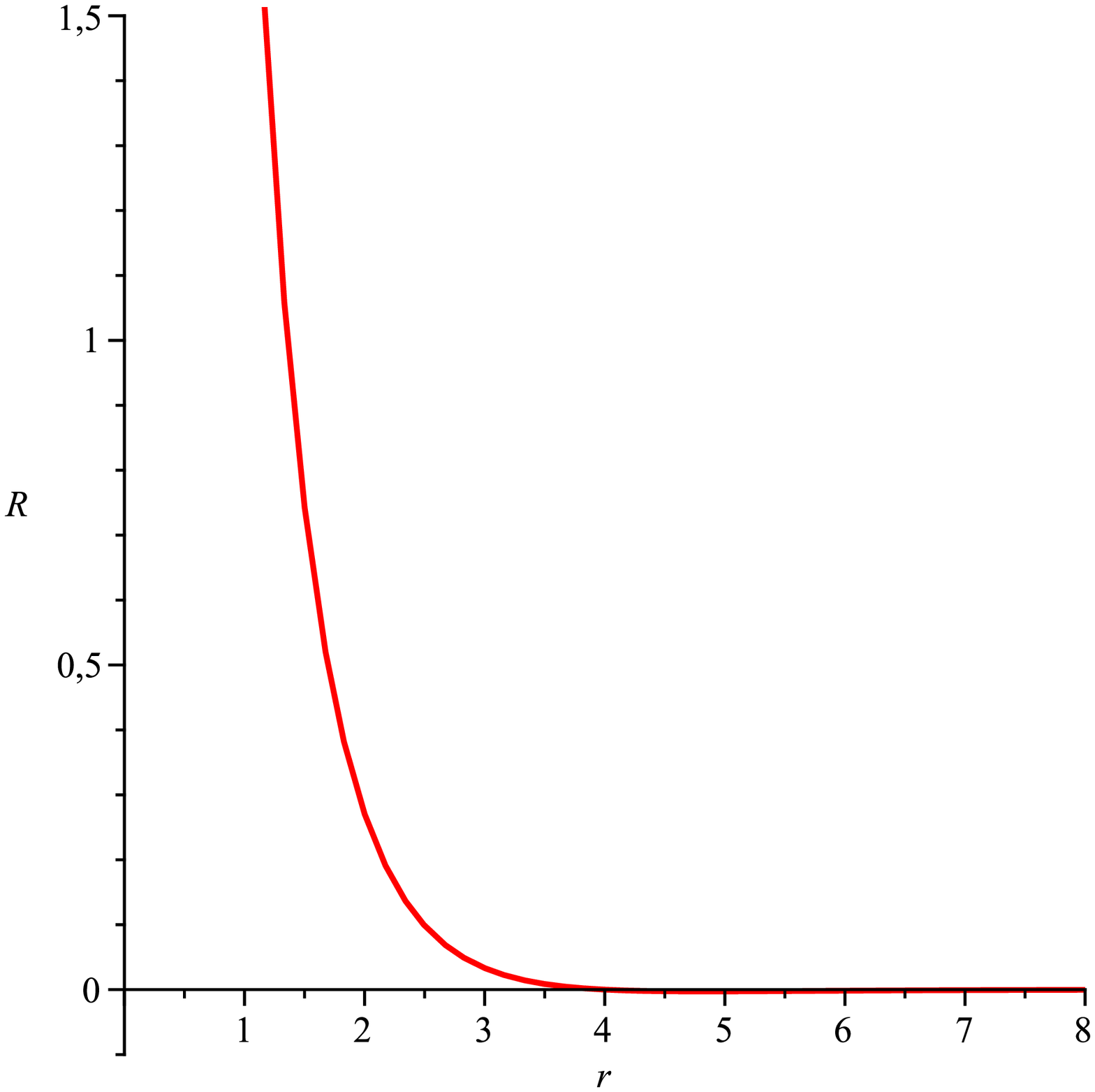}
    \includegraphics[height=7.0cm]{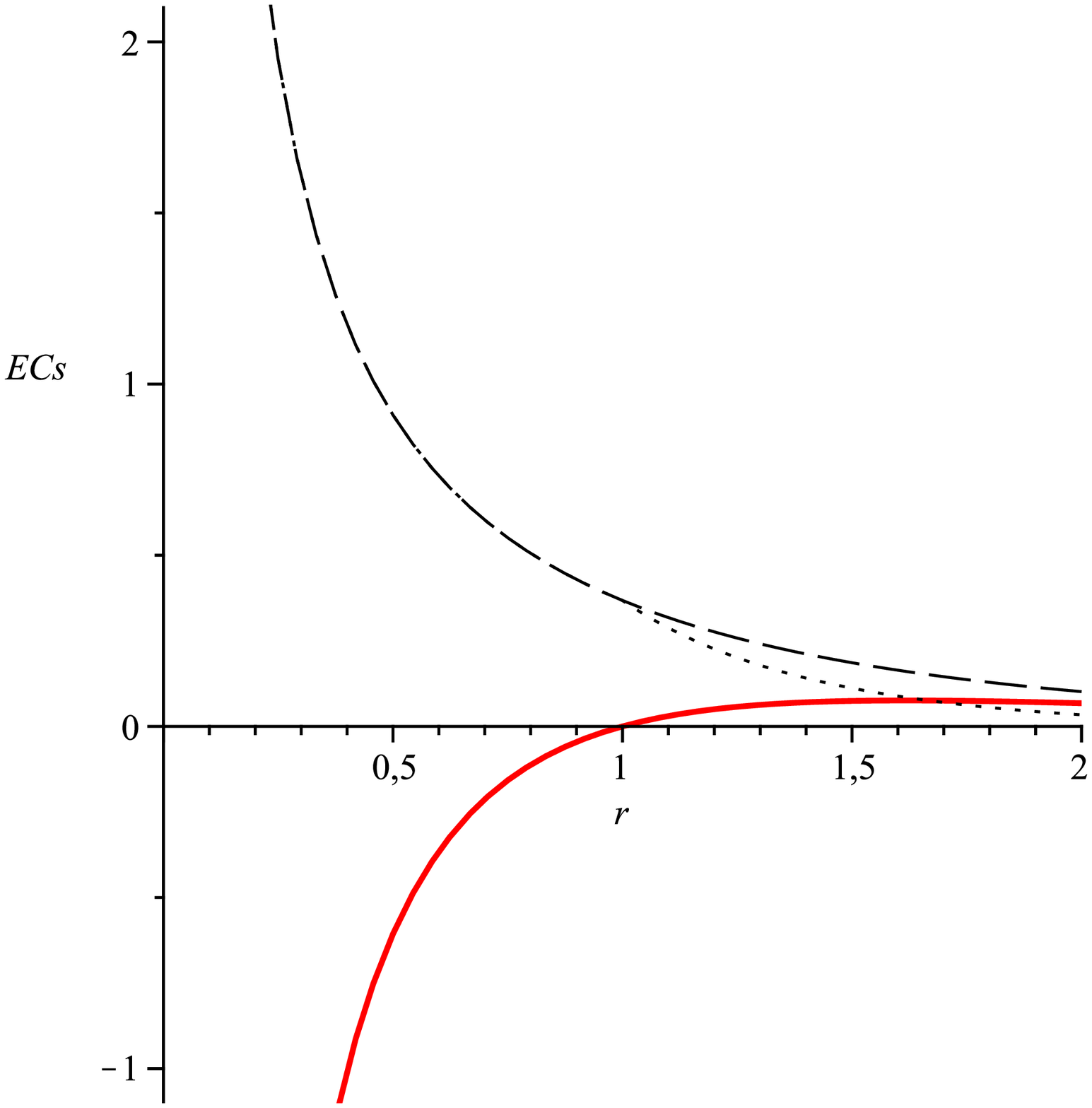}
  \hspace{0.2cm}
      \caption{\label{figG_GUP} On the left we have the Ricci scalar $R$, while on the right the solid curve is the function $(\rho +p_{\mathrm{rad}}+2p_\perp)$ (strong energy condition), the dotted curve is the fucntion $(\rho-|p_\perp|)$ (dominant energy condition) and the dashed curve is $(\rho +p_\perp)$ (null energy condition). In both plots we set $\sqrt{\beta}=1$.}
     \end{center}
  \end{figure}

Here we want to do a step forward by deriving a black hole metric corresponding to a source term affected by the cut off $\sqrt{\beta}$. To achieve this goal we just need to follow the above procedure: we choose a suitable entire function
to model the suppression of higher momenta \textit{\`a la} GUP, {\it i.e.},
\begin{equation}
{\cal A}(p^2)=\left(1+\beta p^2\right)^{1/2},
\end{equation}
where the GUP cut off is $\beta=\ell^2$. The above entire function is of order lower than $1/2$, precluding the possibility of convergent perturbation theories. We will understand later what are the repercussions of such a low order on black hole metrics. 
As a start we need to represent the above function in terms of the D'Alembertian operator. Since our goal is to determine the energy momentum tensor (\ref{exoticstress}), it is more convenient to display the representation of ${\cal A}^{-2}$ instead of ${\cal A}$. By means of a Schwinger representation we write
\begin{equation}
{\cal A}^{-2}\left(\Box\right)=\int_0^\infty ds \ e^{-s\left(1+(-\Box)\right)}.
\end{equation}
Again by employing free falling Cartesian like coordinates we can derive the action of the above operator on the static source, {\it i.e.},
\begin{eqnarray}
{\cal A}^{-2}\left(\Box\right)\delta(\vec{x}) &=&\int_0^\infty ds \ e^{-s\left(1+(-\nabla^2)\right)}
\frac{1}{(2\pi)^3} \int d^3p \ e^{i\vec{x}\cdot\vec{p}} \nonumber \\
&=&
\frac{1}{(2\pi)^3} \int_0^\infty ds \ e^{-s} \int d^3p \ e^{-s\beta \vec{p} ^2} e^{i\vec{x}\cdot\vec{p}}
\nonumber \\
&=&
\frac{1}{(2\pi)^3} \int \frac{d^3 p}{\left(1+\beta \vec{p} ^2\right)}  e^{i\vec{x}\cdot\vec{p}}.
\end{eqnarray}
The above integral is a well known one. It corresponds to the Fourier transform of the Yukawa potential due to the exchange of virtual scalars of mass $\sqrt{\beta}^{-1}$. As a result we have that the energy density reads
\begin{equation}
{\mathfrak T}^0\,_0= - M {\cal A}^{-2}\left(\Box\right)\delta(\vec{x})= -\frac{M}{\beta} \ \frac{e^{-|\vec{x}|/\sqrt{\beta}}}{4\pi|\vec{x}|}.
\label{stressgup}
\end{equation}
A first consequence of working with entire functions of order lower than $1/2$ can be seen by taking the limit at short scales of the above profile, {\it i.e.} $|\vec{x}|\ll \sqrt{\beta}$. The net effect of GUP is just to replace the ``classical'' Dirac delta in (\ref{t00}) with $e^{-|\vec{x}|/\sqrt{\beta}}$, a slightly wider distribution.  However this is not enough: contrary to what happened in the previous case (\ref{enprof}), the cut off $\sqrt{\beta}$ fails to regularize the singularity in the energy density. This fact affects the structure of the spacetime geometry itself. By considering the conservation of the energy momentum tensor, \textit{i.e.}, $\nabla_\mu {\mathfrak T}^{\mu\nu}=0$, one determines the following pressure terms 
\begin{eqnarray}
p_\mathrm{rad}(r)&=& -\frac{M}{\beta} \ \frac{e^{-r/\sqrt{\beta}}}{4\pi r}\\
p_\perp(r)&=&-\frac{M}{2\beta} \ \frac{e^{-r/\sqrt{\beta}}}{4\pi r}+\frac{M}{8\pi\beta^{3/2}} \ e^{-r/\sqrt{\beta}}
\end{eqnarray}
with ${\mathfrak T}_\nu^{\ \mu}=\mathrm{diag}\left(-\rho(r), p_\mathrm{rad}(r), p_\perp(r), p_\perp(r)\right)$ as above. This permits to evaluate the curvature scalar $R=-8\pi G {\mathfrak T}_\mu^{\ \mu}$ by tracing Einstein equations (\ref{e1})
\begin{equation}
R=\left(\frac{8MG}{\beta}\right) \ \frac{e^{-r/\sqrt{\beta}}}{r}-\left(\frac{2MG}{\beta^{3/2}}\right) \ e^{-r/\sqrt{\beta}}.
\end{equation}
The curvature singularity is not removed despite the introduction of the cut off $\sqrt{\beta}$ and the usual local violation of energy conditions (see Fig. (\ref{figG_GUP})). In any case the GUP effects modify the profile of the singularity a fact that discloses some new features.
By integrating (\ref{stressgup}) one determines the profile of the cumulative mass distribution ${\cal M}(r)$ as well as the running coupling ${\cal G}(r)$, \textit{i.e.},
\begin{equation}
{\cal M}(r)/M={\cal G}(r)/G=1-e^{-r/\sqrt{\beta}}-(r/\sqrt{\beta}) e^{-r/\sqrt{\beta}}.
\label{gupcoup}
\end{equation}

   \begin{figure}
 \begin{center}
  \includegraphics[height=7.0cm]{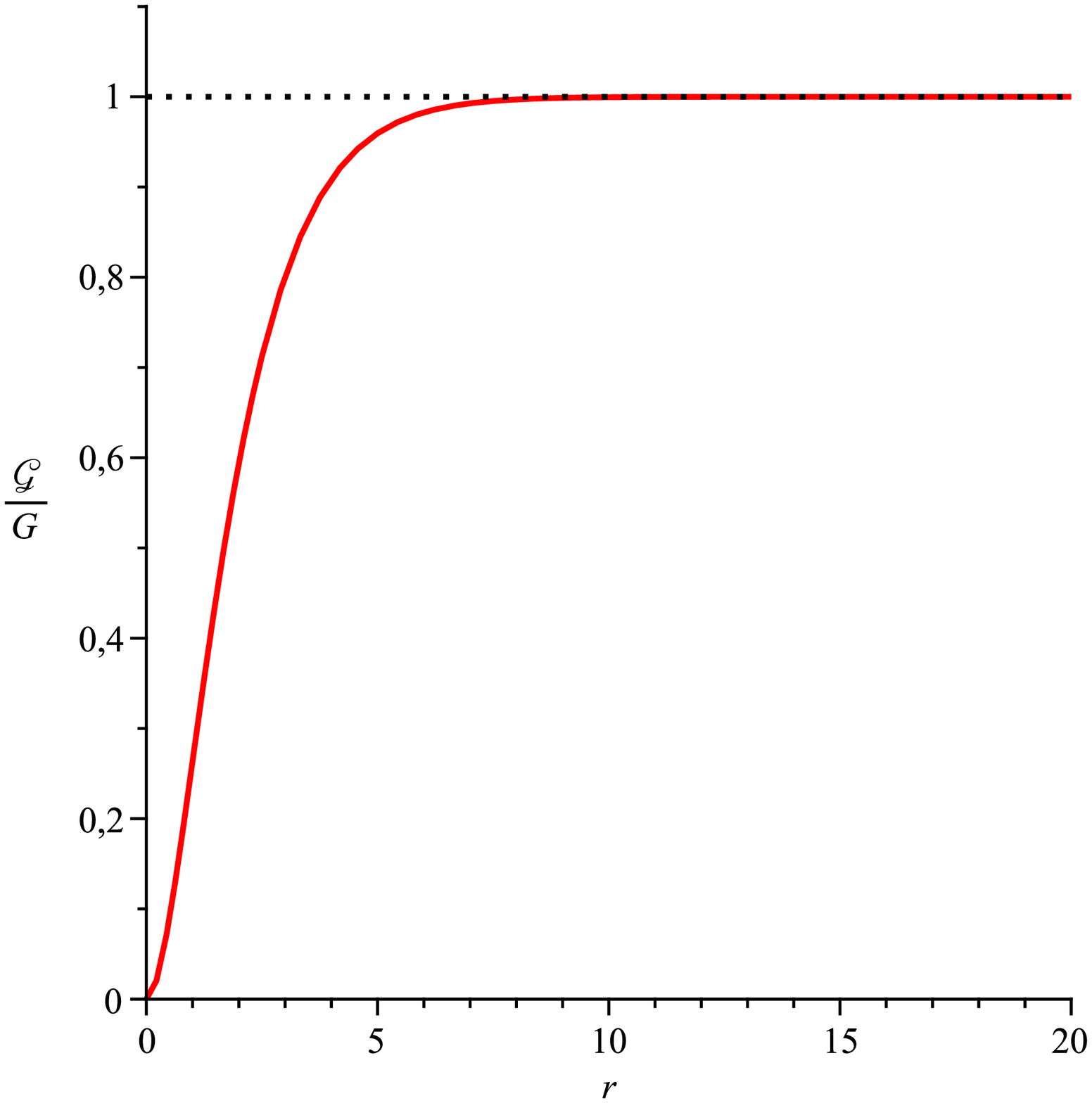}
  \includegraphics[height=7.0cm]{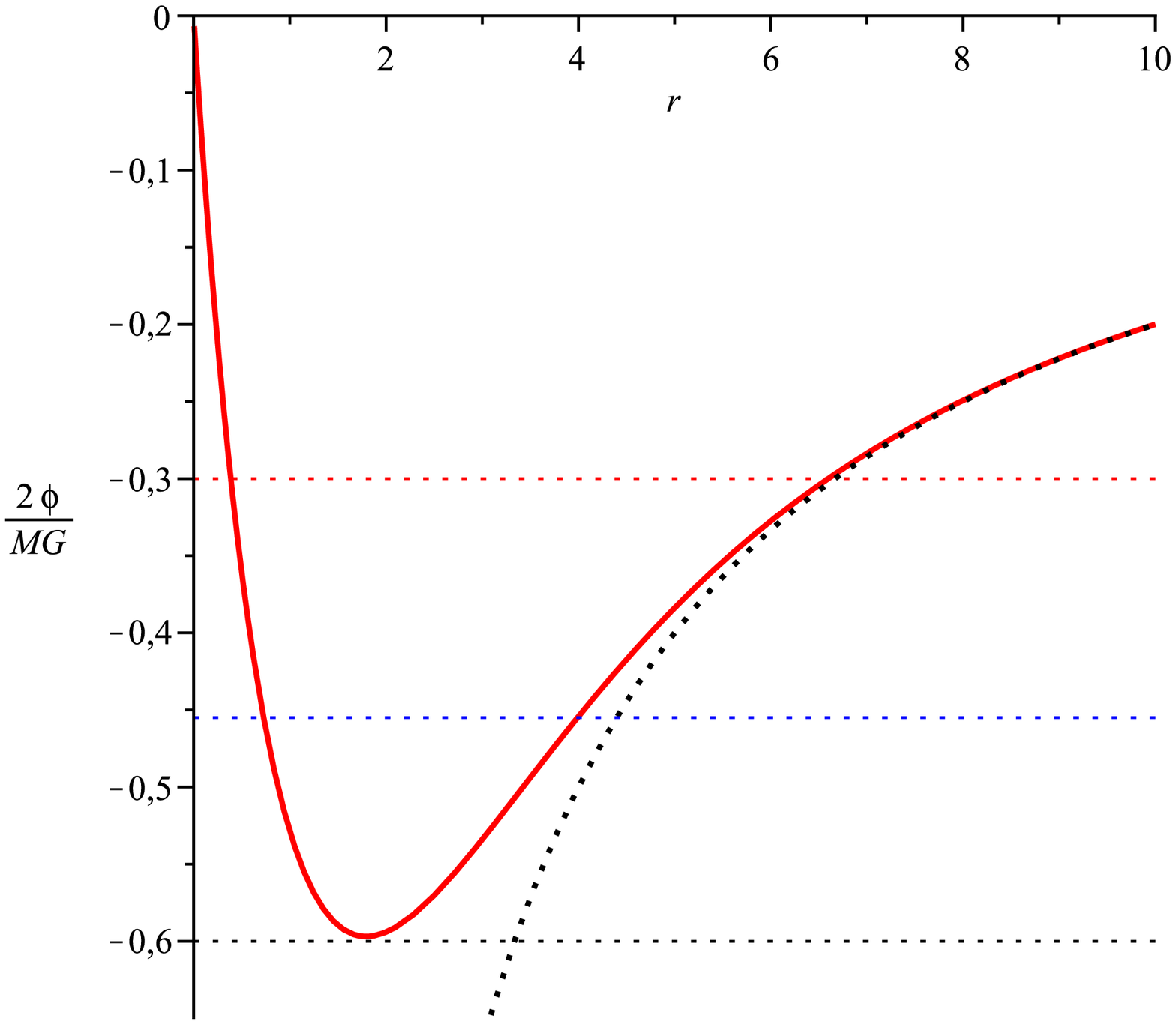}
 \hspace{0.2cm}
      \caption{\label{figphi_GUP} On the left we have the function ${\cal G}(r)/G$ (solid line), while on the right we have the function $y=2\phi (r)/MG=-2{\cal G}(r)/(Gr)$ (solid line)  plotted in comparison with the classical curve $y=-2/r$, by setting $\sqrt{\beta}=1$. Intersections with horizontal curves $y=-1/MG$ give horizon radii. 
 }
 \end{center}
  \end{figure}

In Fig. (\ref{figphi_GUP}), we see that the coupling constant becomes weaker at short scales even if not enough to improve the manifold. Correspondingly ${\cal M}(r)$ is not enough spread out to cure pathologies. In any case at the level of Newtonian potential, the usual $-1/r$ singularity is tamed: by considering the expansion at short scales of $\phi=-G{\cal M}(r)/r$ (or equivalently $-{\cal G}(r)M/r$) we find
\begin{equation}
\phi(r)\approx -GM\left(\frac{r}{\beta}\right), \ \quad r\lesssim \sqrt{\beta}
\end{equation}
while GUP corrections quickly die off at scales $r\gg\sqrt{\beta}$. This behavior of the Newtonian potential has important consequences on the thermodynamic side. We recall that zeros of the Hawking temperature coincide with stationary points of the Newtonian potential. Falling off to zero at both $r\to 0$ and $r\to\infty$, the function $\phi$ must admit a maximum a fact that permits to conclude that the black hole will not evaporate off but will rather condensate into a zero temperature remnant. This can be viewed also in Fig. (\ref{figphi_GUP}): at $r_0\approx 1.73\sqrt{\beta}$ there exists a lower bound $M_0\approx 1.68\sqrt{\beta}/G$ for the mass spectrum below which no horizons form. We conclude that the global structure of the resulting solution
\begin{equation}
ds^2=-\left(1-GM\gamma(2;r/\sqrt{\beta})/r\right)dt^2-\left(1-GM\gamma(2;r/\sqrt{\beta})/r\right)^{-1}dr^2+r^2d\Omega^2
\label{gupspace}
\end{equation}
is analogue of the Reissner-Nordstr\"{o}m geometry with the number of horizons determined by the value of the mass parameter $M$ with respect to $M_0$. Here $\gamma(2;r/\sqrt{\beta})$, the incomplete gamma function
defined by
\begin{equation}
\gamma\left(s; x\right)=\int_0^x\frac{dt}{t}\ t^s \ e^{-t},
\end{equation}
provides an equivalent representation of the function ${\cal G}(r)/G$ in (\ref{gupcoup}). We notice that, even in the absence of charge, the solution develops a naked singularity for $M<M_0$.

\begin{figure}
 \begin{center}
 \includegraphics[height=7.0cm]{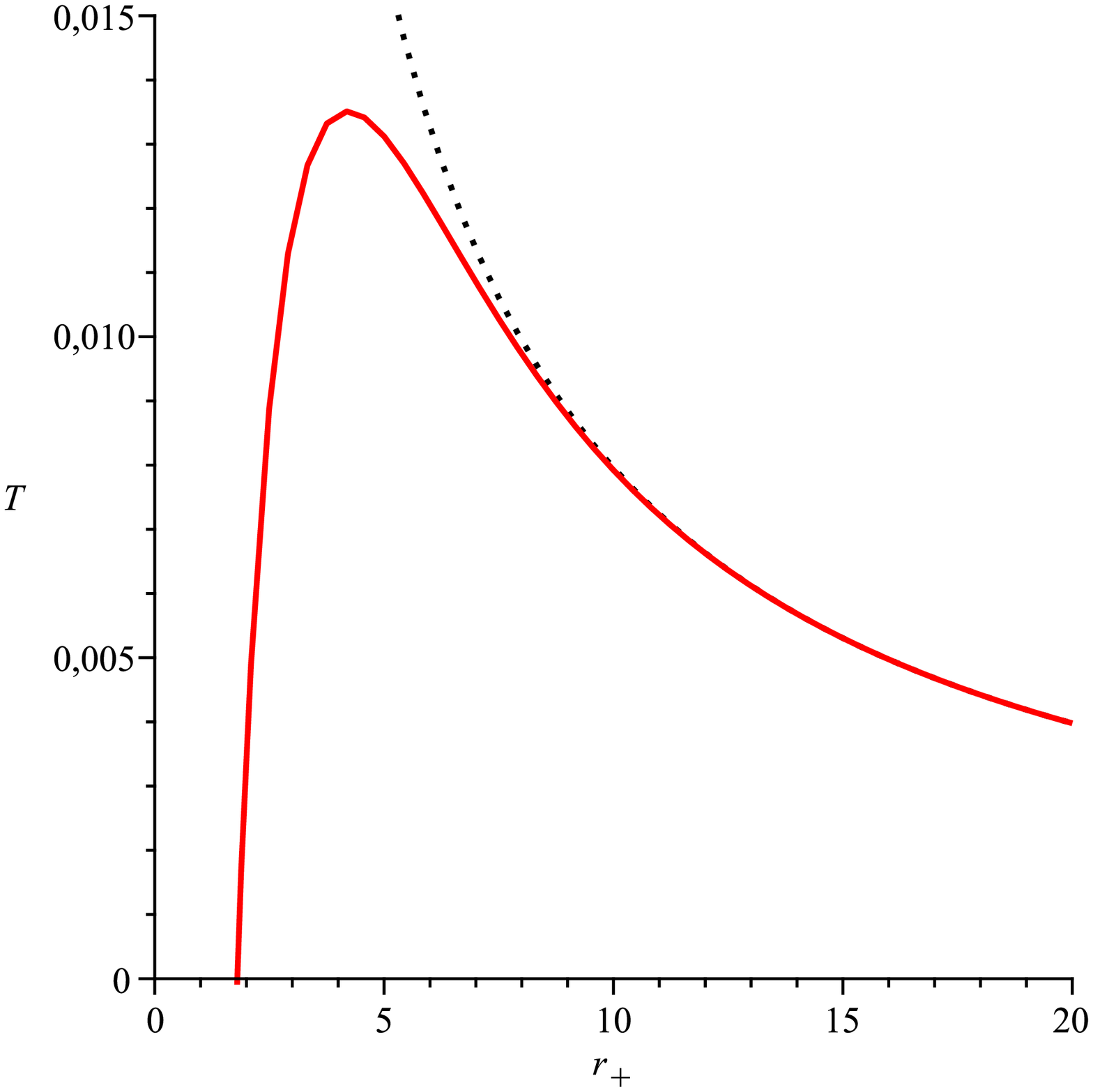}
 \includegraphics[height=7.0cm]{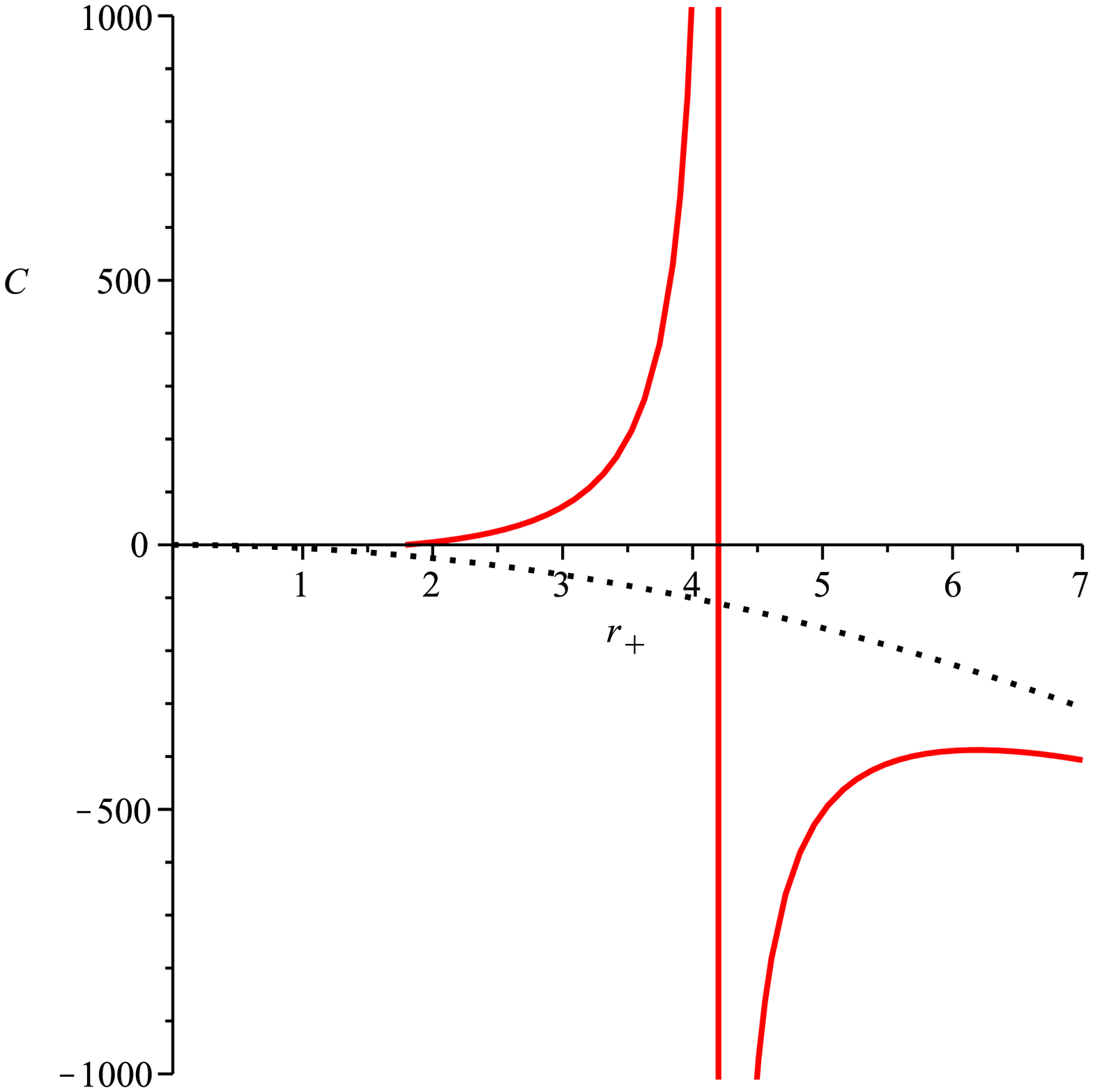}
 \hspace{0.2cm}
      \caption{\label{figtemp_GUP} On the left we have the black hole temperature and on the right its heat capacity. In both plots solid lines represent the new behaviors while dotted lines are the classical results. In addition we set $\sqrt{\beta}=1$.
 }
 \end{center}
  \end{figure}

By following the procedure of the previous section we now display the black hole temperature
\begin{equation}
T=\frac{1}{4\pi r_+}\left(1-\frac{r_+^2}{\beta} \frac{ e^{-r_+/\sqrt{\beta} }}{\gamma(2;r/\sqrt{\beta})} \right).
\end{equation}
Being $T$ vanishing $r_+\to\infty$ as well as $r_+\to r_0$ it must admit a maximum, as is evident in Fig. (\ref{figtemp_GUP}).
Qualitatively the temperature has a behavior in agreement to what found in \cite{APS01,ChA03} for $r_+\geq r_{\mathrm{max}}\approx4.20 \sqrt{\beta}$ with $T_{\mathrm{max}}\equiv T(r_{\mathrm{max}})\approx 1.89\times 10^{-2} \sqrt{\beta}/G$. Our result is a little bit more sophisticated: by recalling that stationary points in the temperature correspond to asymptotes in the heat capacity we obtain a phase transition to from a Schwarzschild phase ($C<0$) to a cooling down phase ($C>0$) with final evaporation switching off ($C=T=0$). As in the case of the previous section quantum back reaction effects turn out to be negligible. This can be seen by comparing the energy of the emitted particles, \textit{i.e.} $T$, with the black hole mass $M$: the ratio $T/M<T_{\mathrm{max}}/M_0\approx 1.13 \times 10^{-2}$ is so small that the spacetime (\ref{gupspace}) remains unmodified during all the evaporation process. 

To confirm the above reasoning we need to determine the black hole heat capacity. As a preliminary step we display the black hole entropy by using the general equation (\ref{BHE})
\begin{equation}
S=\frac{2\pi}{G}\int_{r_0}^{r_+}dr_H\frac{r_H}{\gamma(2; r_H/\sqrt{\beta})}.
\label{sgup}
\end{equation}
Even if we cannot integrate the above formula analytically, it is not hard to prove that at the leading order for $(r_H/\sqrt{\beta})>1$ the entropy fulfills the area law, \textit{i.e.}, $S\approx \pi r^2_+/G\gamma(2; r/\sqrt{\beta})$. On the ground of (\ref{sgup}) we can use (\ref{heatcap1}) to determine the heat capacity that reads
\begin{equation}
C=\frac{8\pi^2 r^2_+}{G\gamma(2;r_+/\sqrt{\beta})}\ T\ \left(H(r_+)-4\pi T\right)^{-1},
\end{equation}
where the function $H(r_+)$ is defined by
\begin{equation}
H(r_+)\equiv-\frac{\gamma^{\prime}(2; r_+/\sqrt{\beta})}{\gamma(2; r_+/\sqrt{\beta})}-r_+\frac{\gamma^{\prime\prime}(2; r_+/\sqrt{\beta})}{\gamma(2; r_+/\sqrt{\beta})}+r_+\left(\frac{\gamma^{\prime}(2; r_+/\sqrt{\beta})}{\gamma(2; r_+/\sqrt{\beta})}\right)^2,
\end{equation}
with the derivatives of the gamma functions
\begin{eqnarray}
\gamma^{\prime}(2; r_+/\sqrt{\beta})&=&\frac{r_+}{\beta}e^{-r_+/\sqrt{\beta}}\\
\gamma^{\prime\prime}(2; r_+/\sqrt{\beta})&=&\frac{1}{\beta}e^{-r_+/\sqrt{\beta}}\left(1-\frac{r_+}{\sqrt{\beta}}\right).
\end{eqnarray} 
The term $\left(H(r_+)-4\pi T\right)=4\pi r_+\left(dT/dr_+\right)$ determines the sign of $C$ according to our expectations. In particular when $\left(dT/dr_+\right)=0$ (i.e. at $r=r_{\mathrm{max}}$) the heat capacity has a vertical asymptote and the black hole undergoes a phase transition. The resulting plot is given in Fig. \ref{figtemp_GUP}.

\section{Conclusions}
\label{quattro}

In this paper we have analyzed the possibility for nonlocality to be the key ingredient for curing the conventional pathologies of classical gravity at short scales. To reach this goal, we have derived black hole solutions by extending our studies on nonlocal gravitational actions with entire function of order $\alpha>1/2$ \cite{NSS06b,NiS10,MMN11} to the case of entire functions of order $\alpha\leq 1/2$. Specifically we have derived a black hole solution corresponding to the case $\alpha=1/2$ and another one corresponding to GUP corrections modeled in terms of an entire function belonging to the class $\alpha<1/2$.   We have now a more complete scenario and we are ready to summarize our results by pinpointing the general features of these black holes.

   \begin{table}[pht!]
 {\begin{tabular}{@{}c|ccc|c|c|c|r|c|c|c@{}} 
  &  & $r_0 [\ell]$ & & $M_0 [\ell/G]$ &  $r_{\mathrm{max}}  [\ell]$ &  $T_{\mathrm{max}}  [\ell/G]$  & $(T/M)<\ \ $ & Newton $\phi$ & curvature & thermodynamics \\
 \colrule ``NC-Schw'' & & $3.02$  & & $1.90$  & $4.76$  & $0.0150$ & $7.89\times 10^{-3}$ & regular & regular & locally stable
  \\  
  \colrule ``NC-Dirty'' & & $3.02$ & & $1.90$  & $4.84$  & $0.0148$ & $7.79\times 10^{-3}$ & regular & regular &locally stable
  \\  
  \colrule ``$\alpha=1/2$'' & & $1.83$ & & $2.21$  & $3.60$  & $0.0119$ & $5.38\times 10^{-3}$ & regular & regular & locally stable
   \\ 
 \colrule ``GUP''& & $1.73$ & & $1.68$  & $4.20$  & $0.0189$ & $11.25\times 10^{-3}$ & regular & singular & locally stable
    
\end{tabular} 
\caption{Summary of parameters and properties of some families of short scale improved black hole metrics. Specifically we have the noncommutative Schwarzschild black hole (``NC-Schw'') \cite{NSS06b,MMN11},  the noncommutative dirty black hole (``NC-Dirty'') \cite{NiS10} - both corresponding to nonlocal gravity solutions with entire function of order higher than $1/2$ - the nonlocal gravity black hole with entire function of order $1/2$ (``$\alpha=1/2$'') and the GUP modified black hole (``GUP'') - corresponding to a nonlocal gravity solution with entire function of order lower than $1/2$.}
\label{tab1}}
\end{table}

As expected from the general theory \cite{Efi67}, nonlocality determines a UV complete theory of gravity only if the order of entire functions is $\alpha\geq 1/2$. We showed that this can be translated into the ``language of black holes'' by saying that curvature singularities are replaced by  regular deSitter cores which account for the quantum vacuum energy of virtual gravitons. In the case of  $\alpha< 1/2$ such a vacuum energy develops (a fact that can be viewed by looking at the violation of energy conditions in the vicinity of the origin) but it is not regularized. The net effect is just a deformation of the profile of the curvature singularity whose degree of divergence is decreased. As a result for all $\alpha$ one finds that the Newtonian potential admits a maximum and vanishes at the origin. On the other hand at large scales nonlocal effects quickly die off and all black hole metrics consistently match the classical Schwarzschild geometry. We showed that the presence of stationary points of the Newtonian potential for all $\alpha$ has repercussions in two directions: i) the global structure of the corresponding black hole solutions is modified by allowing the horizon extremization; ii) the Hawking temperature does not diverge because the black hole undergoes a phase transition from a negative heat capacity evaporation phase to a positive heat capacity cooling down which terminates with a  zero temperature remnant configuration. Accordingly the black hole temperature never reaches values high enough to disrupt the geometry by permitting a consistent semiclassical description of all the evaporation process. The remnant mass is not only the mass of the black hole in the extremal configuration but also the threshold mass for developing horizons: lighter masses lead to horizonless regular geometries for $\alpha\geq 1/2$ and naked singularities for $\alpha<1/2$.  In addition for all $\alpha$ the area law of the black hole entropy is preserved in form: the only modification occurs at the level of coupling constant where the usual $G$ is replaced by the nonlocal running coupling ${\cal G}(r_+)$.

In summary we showed that the way nonlocality deforms the short distance behavior of gravity has model independent characters with the only exception of the singularity avoidance in the case $\alpha<1/2$. This fact opens the route to further considerations and new investigations. Specifically we can extend some recent results obtained for the case $\alpha>1/2$ to the case of generic $\alpha$.
Concerning the phenomenology of microscopic black holes in particle detectors nonlocal deformations of black hole thermodynamics might lead to distinctive signatures. Even if in the present paper we studied the four dimensional case only we argue that the possibility of having a maximum black hole temperature would imply a Hawking emission in terms of soft particle on the brane, a fact that is in marked contrast to the results for Schwarzschild black holes as shown in \cite{NiW11,BNS11}. On the other hand in the presence of a non-asymptotically flat term, \textit{i.e.} a cosmological constant, nonlocal black holes have a thermodynamics equivalent to a Van der Waals gas in which the role of the average volume occupied by a particle is played by the cut off $\ell$ (or equivalently $\sqrt{\beta}$ in the GUP case) \cite{NiT11}. On the ground of this result, one may think to exploit the black hole phase transition to describe, via the gauge-gravity duality paradigm, the nuclear matter transition from the ``classical'' skyrme crystal to a phase where quantum effects weaken the bond between baryons and eventually melt the crystal into a nuclear gas with consequent switch of the deconfinement phase into a cross over (for more details see \cite{LoT11}).
As a related issue to black holes we recall that the UV completeness of these theories can lead to a two-dimensional Planck scale spectral dimension for a diffusive process propagating over a nonlocal background manifold \cite{MoN10a}. This fact is important since it lets us conclude that nonlocality might be the key ingredient to disclose self-renormalizability properties of gravity at the Planck scale, by virtue of a mechanism of dynamical dimensional reduction.

All the above considerations hold at a qualitative level while the families of nonlocal black holes are quantitatively slightly different (see Tab. \ref{tab1} for a summary of their properties and parameters).
We think that their differences are an advantage rather than a limitation. For instance having different values for the threshold mass implies a variety of scenarios for the production of black holes, that, by considering only the case of $\alpha>1/2$ theories, would be out of reach of the present ground based experiments \cite{MNS11}. In an ideal world one would like to reverse the logic by trying to determine the exact profile of the entire function ${\cal A}$ from experiments. This can be done also in the absence of extradimension: it has been shown that neutrino oscillations are sensible to nonlocal gravity modifications opening the possibility of direct observations in neutrino telescopes like IceCube and ANTARES \cite{SNB11a}.

We conclude by saying that nonlocal deformations of gravity offer reliable solutions to long standing problems in quantum gravity and a rich phenomenology with distinctive predictions that might be potentially observable in a near future.

 \begin{acknowledgements}
This work is supported by the Helmholtz International Center for FAIR within the
framework of the LOEWE program (Landesoffensive zur Entwicklung Wissenschaftlich-\"{O}konomischer
Exzellenz) launched by the State of Hesse and partially by the European Cooperation in Science and Technology (COST) action MP0905 ``Black Holes in a Violent Universe''. The author is grateful to J. Moffat for a careful reading of an early version of the manuscript.
\end{acknowledgements}

\appendix
\section{Useful formulas}
The expansion of the entropy (\ref{entrapp}) in powers of $x\equiv r_H/\ell>1$ can be obtained by the following relation
\begin{eqnarray}
\frac{{\cal G}^\prime(x)}{{\cal G}^2(x)} A_{H}(x)
&=&{\pi }^{2}{x}^{2} \left( 2\, \left( {x}^{2}+1 \right) ^{-1}-2\,{\frac 
{1}{{x}^{2} \left( 1+{x}^{-2} \right) }}+4\,{\frac {1}{{x}^{4} \left( 
1+{x}^{-2} \right) ^{2}}} \right)  \left( \arctan \left( x \right) +{
\frac {1}{x \left( 1+{x}^{-2} \right) }} \right) ^{-2} \nonumber \\
&\approx & 16\,{x}^{-2}-32\,{x}^{-4}+{\frac {128}{3}}\,{\frac {1}{\pi \,{x}^{5}}}
+48\,{x}^{-6}-{\frac {2048}{15}}\,{\frac {1}{\pi \,{x}^{7}}}+O \left( 
{x}^{-8} \right). 
\label{gexp}
\end{eqnarray}
\section{Black hole heat capacity}
The heat capacity in (\ref{heatcap1}) reads
\begin{eqnarray}\label{heatcap}
&& C (r_+)= \left( \frac{2\ell^2 r_+ T}{\pi {\cal G}(r_+)} \right)\times\\ && \left\{ \frac{1}{4\pi (r_+/\ell)}\, \left( \frac{\pi {\cal G}(r_+)}{2G}
 \right)\left[ \frac{-6\,{(r_+/\ell)}^{2}}{ \left( 1+{(r_+/\ell)}^{2}
 \right) ^{2} }+ \frac {8\,{(r_+/\ell)}^{4}}{ \left( 1+{(r_+/\ell)}^{2} \right) ^{3}} 
+\frac{4\,{(r_+/\ell)}^
{5}}{ \left( 1+{(r_+/\ell)}^{2} \right) ^{4}} \left(\frac{ \pi {\cal G}(r_+)}{2G} \right)  \right]
\right. \nonumber
\left. 
 -\frac{1}{4\pi (r_+/\ell)^2}\,
 \left( 4\pi r_+ T \right)  \right\} ^{-1} \nonumber
\end{eqnarray}


\begin{thebibliography}{62}
\expandafter\ifx\csname natexlab\endcsname\relax\def\natexlab#1{#1}\fi
\expandafter\ifx\csname bibnamefont\endcsname\relax
  \def\bibnamefont#1{#1}\fi
\expandafter\ifx\csname bibfnamefont\endcsname\relax
  \def\bibfnamefont#1{#1}\fi
\expandafter\ifx\csname citenamefont\endcsname\relax
  \def\citenamefont#1{#1}\fi
\expandafter\ifx\csname url\endcsname\relax
  \def\url#1{\texttt{#1}}\fi
\expandafter\ifx\csname urlprefix\endcsname\relax\def\urlprefix{URL }\fi
\providecommand{\bibinfo}[2]{#2}
\providecommand{\eprint}[2][]{\url{#2}}

\bibitem[{\citenamefont{Heisenberg}(1938)}]{Hei38}
\bibinfo{author}{\bibfnamefont{W.}~\bibnamefont{Heisenberg}},
  \bibinfo{journal}{Ann. Phys.} \textbf{\bibinfo{volume}{32}},
  \bibinfo{pages}{20} (\bibinfo{year}{1938}), \bibinfo{note}{in *Miller, A.I.:
  \textit{Early quantum electrodynamics}* 244-253 (1996).}

\bibinfo{author}{\bibfnamefont{H.~S.} \bibnamefont{Snyder}},
  \bibinfo{journal}{Phys. Rev.} \textbf{\bibinfo{volume}{71}},
  \bibinfo{pages}{38} (\bibinfo{year}{1947}{\natexlab{a}}).

\bibinfo{author}{\bibfnamefont{H.~S.} \bibnamefont{Snyder}},
  \bibinfo{journal}{Phys. Rev.} \textbf{\bibinfo{volume}{72}},
  \bibinfo{pages}{68} (\bibinfo{year}{1947}{\natexlab{b}}).

\bibinfo{author}{\bibfnamefont{C.~N.} \bibnamefont{Yang}},
  \bibinfo{journal}{Phys. Rev.} \textbf{\bibinfo{volume}{72}},
  \bibinfo{pages}{874} (\bibinfo{year}{1947}).

\bibinfo{author}{\bibfnamefont{G.}~\bibnamefont{Veneziano}},
  \bibinfo{journal}{Nuovo. Cim.} \textbf{\bibinfo{volume}{A57}},
  \bibinfo{pages}{190} (\bibinfo{year}{1968}).

\bibitem[{\citenamefont{Polyakov}(1981)}]{Pol81}
\bibinfo{author}{\bibfnamefont{A.~M.} \bibnamefont{Polyakov}},
  \bibinfo{journal}{Phys. Lett.} \textbf{\bibinfo{volume}{B103}},
  \bibinfo{pages}{207} (\bibinfo{year}{1981}).

\bibinfo{author}{\bibfnamefont{R.~J.} \bibnamefont{Riegert}},
  \bibinfo{journal}{Phys. Lett.} \textbf{\bibinfo{volume}{B134}},
  \bibinfo{pages}{56} (\bibinfo{year}{1984}).
  
   J.~W.~Moffat,
  Phys.\ Rev.\  D {\bf 39}, 3654 (1989).

   J.~W.~Moffat,
  Phys.\ Rev.\  D {\bf 41}, 1177 (1990).

   D.~Evens, J.~W.~Moffat, G.~Kleppe and R.~P.~Woodard,
  Phys.\ Rev.\  D {\bf 43}, 499 (1991).


\bibinfo{author}{\bibfnamefont{I.~Y.} \bibnamefont{Aref'eva}} \bibnamefont{and}
  \bibinfo{author}{\bibfnamefont{L.~V.} \bibnamefont{Joukovskaya}},
  \bibinfo{journal}{JHEP} \textbf{\bibinfo{volume}{10}}, \bibinfo{pages}{087}
  (\bibinfo{year}{2005}). 

\bibinfo{author}{\bibfnamefont{T.}~\bibnamefont{Biswas}},
  \bibinfo{author}{\bibfnamefont{A.}~\bibnamefont{Mazumdar}}, \bibnamefont{and}
  \bibinfo{author}{\bibfnamefont{W.}~\bibnamefont{Siegel}},
  \bibinfo{journal}{JCAP} \textbf{\bibinfo{volume}{0603}}, \bibinfo{pages}{009}
  (\bibinfo{year}{2006}). 

  E.~Spallucci, A.~Smailagic and P.~Nicolini,
  Phys.\ Rev.\  D {\bf 73}, 084004 (2006).
  
\bibinfo{author}{\bibfnamefont{S.}~\bibnamefont{Deser}} \bibnamefont{and}
  \bibinfo{author}{\bibfnamefont{R.~P.} \bibnamefont{Woodard}},
  \bibinfo{journal}{Phys. Rev. Lett.} \textbf{\bibinfo{volume}{99}},
  \bibinfo{pages}{111301} (\bibinfo{year}{2007}). 

\bibinfo{author}{\bibfnamefont{H.-J.} \bibnamefont{Blome}},
  \bibinfo{author}{\bibfnamefont{C.}~\bibnamefont{Chicone}},
  \bibinfo{author}{\bibfnamefont{F.~W.} \bibnamefont{Hehl}}, \bibnamefont{and}
  \bibinfo{author}{\bibfnamefont{B.}~\bibnamefont{Mashhoon}},
  \bibinfo{journal}{Phys. Rev.} \textbf{\bibinfo{volume}{D81}},
  \bibinfo{pages}{065020} (\bibinfo{year}{2010}). 

\bibinfo{author}{\bibfnamefont{G.}~\bibnamefont{Calcagni}} \bibnamefont{and}
  \bibinfo{author}{\bibfnamefont{G.}~\bibnamefont{Nardelli}},
  \bibinfo{journal}{Phys. Rev.} \textbf{\bibinfo{volume}{D82}},
  \bibinfo{pages}{123518} (\bibinfo{year}{2010}). 
  
  T.~Biswas, T.~Koivisto and A.~Mazumdar,
  JCAP {\bf 1011}, 008 (2010).

  J.~W.~Moffat,
  Eur.\ Phys.\ J.\ Plus {\bf 126}, 53 (2011).

\bibinfo{author}{\bibfnamefont{T.}~\bibnamefont{Biswas}},
  \bibinfo{author}{\bibfnamefont{E.}~\bibnamefont{Gerwick}},
  \bibinfo{author}{\bibfnamefont{T.}~\bibnamefont{Koivisto}}, \bibnamefont{and}
  \bibinfo{author}{\bibfnamefont{A.}~\bibnamefont{Mazumdar}},
  \bibinfo{journal}{Phys.Rev.Lett.} \textbf{\bibinfo{volume}{108}},
  \bibinfo{pages}{031101} (\bibinfo{year}{2012}). 
  
 


\bibitem[{\citenamefont{Gaete et~al.}(2010)\citenamefont{Gaete, Helayel-Neto,
  and Spallucci}}]{GHS10}
\bibinfo{author}{\bibfnamefont{P.}~\bibnamefont{Gaete}},
  \bibinfo{author}{\bibfnamefont{J.~A.} \bibnamefont{Helayel-Neto}},
  \bibnamefont{and}
  \bibinfo{author}{\bibfnamefont{E.}~\bibnamefont{Spallucci}},
  \bibinfo{journal}{Phys. Lett.} \textbf{\bibinfo{volume}{B693}},
  \bibinfo{pages}{155} (\bibinfo{year}{2010}). 

\bibinfo{author}{\bibfnamefont{J.~R.} \bibnamefont{Mureika}} \bibnamefont{and}
  \bibinfo{author}{\bibfnamefont{E.}~\bibnamefont{Spallucci}},
  \bibinfo{journal}{Phys. Lett.} \textbf{\bibinfo{volume}{B693}},
  \bibinfo{pages}{129} (\bibinfo{year}{2010}). 

\bibitem[{\citenamefont{Moffat}(2011)}]{Mof10}
\bibinfo{author}{\bibfnamefont{J.~W.} \bibnamefont{Moffat}},
  \bibinfo{journal}{Eur. Phys. J. Plus} \textbf{\bibinfo{volume}{126}},
  \bibinfo{pages}{43} (\bibinfo{year}{2011}). 

\bibitem[{\citenamefont{Modesto et~al.}(2011)\citenamefont{Modesto, Moffat, and
  Nicolini}}]{MMN11}
\bibinfo{author}{\bibfnamefont{L.}~\bibnamefont{Modesto}},
  \bibinfo{author}{\bibfnamefont{J.~W.} \bibnamefont{Moffat}},
  \bibnamefont{and} \bibinfo{author}{\bibfnamefont{P.}~\bibnamefont{Nicolini}},
  \bibinfo{journal}{Phys. Lett.} \textbf{\bibinfo{volume}{B695}},
  \bibinfo{pages}{397} (\bibinfo{year}{2011}). 
 
\bibitem[{\citenamefont{Modesto}(2011)}]{Mod11}
\bibinfo{author}{\bibfnamefont{L.}~\bibnamefont{Modesto}}
  (\bibinfo{year}{2011}), \eprint{arXiv:1107.2403}.

\bibinfo{author}{\bibfnamefont{L.}~\bibnamefont{Modesto}}
  (\bibinfo{year}{2012}), \eprint{arXiv:1202.0008}.

\bibitem[{\citenamefont{Efimov}(1967)}]{Efi67}
\bibinfo{author}{\bibfnamefont{G.~V.} \bibnamefont{Efimov}},
  \bibinfo{journal}{Commun. Math. Phys.} \textbf{\bibinfo{volume}{5}},
  \bibinfo{pages}{42} (\bibinfo{year}{1967}).

\bibinfo{author}{\bibfnamefont{G.~V.} \bibnamefont{Efimov}},
  \emph{\bibinfo{title}{Nonlocal interactions of quantum fields}}
  (\bibinfo{publisher}{Nauka Publishers}, \bibinfo{address}{Moscow, Russia},
  \bibinfo{year}{1977}).

\bibinfo{author}{\bibfnamefont{G.~V.} \bibnamefont{Efimov}},
  \emph{\bibinfo{title}{Problems of the quantum theory of nonlocal
  interactions}} (\bibinfo{publisher}{Nauka Publishers},
  \bibinfo{address}{Moscow, Russia}, \bibinfo{year}{1985}).


\bibitem[{\citenamefont{Nicolini}(2005)}]{Nic05}
\bibinfo{author}{\bibfnamefont{P.}~\bibnamefont{Nicolini}},
  \bibinfo{journal}{J. Phys.} \textbf{\bibinfo{volume}{A38}},
  \bibinfo{pages}{L631} (\bibinfo{year}{2005}). 

\bibinfo{author}{\bibfnamefont{P.}~\bibnamefont{Nicolini}},
  \bibinfo{author}{\bibfnamefont{A.}~\bibnamefont{Smailagic}},
  \bibnamefont{and}
  \bibinfo{author}{\bibfnamefont{E.}~\bibnamefont{Spallucci}},
  \bibinfo{journal}{ESA Spec. Publ.} \textbf{\bibinfo{volume}{637}},
  \bibinfo{pages}{11.1} (\bibinfo{year}{2006}{\natexlab{a}}).



\bibinfo{author}{\bibfnamefont{T.~G.} \bibnamefont{Rizzo}},
  \bibinfo{journal}{JHEP} \textbf{\bibinfo{volume}{09}}, \bibinfo{pages}{021}
  (\bibinfo{year}{2006}). 

\bibinfo{author}{\bibfnamefont{S.}~\bibnamefont{Ansoldi}},
  \bibinfo{author}{\bibfnamefont{P.}~\bibnamefont{Nicolini}},
  \bibinfo{author}{\bibfnamefont{A.}~\bibnamefont{Smailagic}},
  \bibnamefont{and}
  \bibinfo{author}{\bibfnamefont{E.}~\bibnamefont{Spallucci}},
  \bibinfo{journal}{Phys. Lett.} \textbf{\bibinfo{volume}{B645}},
  \bibinfo{pages}{261} (\bibinfo{year}{2007}). 

 R.~Casadio and P.~Nicolini,
  JHEP {\bf 0811}, 072 (2008).
  
\bibinfo{author}{\bibfnamefont{E.}~\bibnamefont{Spallucci}},
  \bibinfo{author}{\bibfnamefont{A.}~\bibnamefont{Smailagic}},
  \bibnamefont{and} \bibinfo{author}{\bibfnamefont{P.}~\bibnamefont{Nicolini}},
  \bibinfo{journal}{Phys. Lett.} \textbf{\bibinfo{volume}{B670}},
  \bibinfo{pages}{449} (\bibinfo{year}{2009}). 


\bibinfo{author}{\bibfnamefont{R.}~\bibnamefont{Banerjee}},
  \bibinfo{author}{\bibfnamefont{S.}~\bibnamefont{Gangopadhyay}},
  \bibnamefont{and} \bibinfo{author}{\bibfnamefont{S.~K.} \bibnamefont{Modak}},
  \bibinfo{journal}{Phys. Lett.} \textbf{\bibinfo{volume}{B686}},
  \bibinfo{pages}{181} (\bibinfo{year}{2010}). 

\bibinfo{author}{\bibfnamefont{D.~M.} \bibnamefont{Gingrich}},
  \bibinfo{journal}{JHEP} \textbf{\bibinfo{volume}{05}}, \bibinfo{pages}{022}
  (\bibinfo{year}{2010}). 

\bibinfo{author}{\bibfnamefont{A.}~\bibnamefont{Smailagic}} \bibnamefont{and}
  \bibinfo{author}{\bibfnamefont{E.}~\bibnamefont{Spallucci}},
  \bibinfo{journal}{Phys. Lett.} \textbf{\bibinfo{volume}{B688}},
  \bibinfo{pages}{82} (\bibinfo{year}{2010}). 

\bibinfo{author}{\bibfnamefont{L.}~\bibnamefont{Modesto}} \bibnamefont{and}
  \bibinfo{author}{\bibfnamefont{P.}~\bibnamefont{Nicolini}},
  \bibinfo{journal}{Phys. Rev.} \textbf{\bibinfo{volume}{D82}},
  \bibinfo{pages}{104035} (\bibinfo{year}{2010}{\natexlab{a}}).

\bibinfo{author}{\bibfnamefont{R.~B.} \bibnamefont{Mann}} \bibnamefont{and}
  \bibinfo{author}{\bibfnamefont{P.}~\bibnamefont{Nicolini}},
  \bibinfo{journal}{Phys. Rev.} \textbf{\bibinfo{volume}{D84}},
  \bibinfo{pages}{064014} (\bibinfo{year}{2011}). 


\bibinfo{author}{\bibfnamefont{J.~R.} \bibnamefont{Mureika}} \bibnamefont{and}
  \bibinfo{author}{\bibfnamefont{P.}~\bibnamefont{Nicolini}},
  \bibinfo{journal}{Phys. Rev.} \textbf{\bibinfo{volume}{D84}},
  \bibinfo{pages}{044020} (\bibinfo{year}{2011}). 

\bibitem[{\citenamefont{Nicolini
  et~al.}(2006{\natexlab{b}})\citenamefont{Nicolini, Smailagic, and
  Spallucci}}]{NSS06b}
\bibinfo{author}{\bibfnamefont{P.}~\bibnamefont{Nicolini}},
  \bibinfo{author}{\bibfnamefont{A.}~\bibnamefont{Smailagic}},
  \bibnamefont{and}
  \bibinfo{author}{\bibfnamefont{E.}~\bibnamefont{Spallucci}},
  \bibinfo{journal}{Phys. Lett.} \textbf{\bibinfo{volume}{B632}},
  \bibinfo{pages}{547} (\bibinfo{year}{2006}{\natexlab{b}}).


\bibitem[{\citenamefont{Nicolini and Spallucci}(2010)}]{NiS10}
\bibinfo{author}{\bibfnamefont{P.}~\bibnamefont{Nicolini}} \bibnamefont{and}
  \bibinfo{author}{\bibfnamefont{E.}~\bibnamefont{Spallucci}},
  \bibinfo{journal}{Class. Quant. Grav.} \textbf{\bibinfo{volume}{27}},
  \bibinfo{pages}{015010} (\bibinfo{year}{2010}). 

\bibitem[{\citenamefont{Nicolini and Torrieri}(2011)}]{NiT11}
\bibinfo{author}{\bibfnamefont{P.}~\bibnamefont{Nicolini}} \bibnamefont{and}
  \bibinfo{author}{\bibfnamefont{G.}~\bibnamefont{Torrieri}},
  \bibinfo{journal}{JHEP} \textbf{\bibinfo{volume}{08}}, \bibinfo{pages}{097}
  (\bibinfo{year}{2011}). 


\bibitem[{\citenamefont{Nicolini}(2009)}]{Nic09}
\bibinfo{author}{\bibfnamefont{P.}~\bibnamefont{Nicolini}},
  \bibinfo{journal}{Int. J. Mod. Phys.} \textbf{\bibinfo{volume}{A24}},
  \bibinfo{pages}{1229} (\bibinfo{year}{2009}). 


\bibitem[{\citenamefont{Banerjee et~al.}(2008)\citenamefont{Banerjee, Majhi,
  and Samanta}}]{BMS08}
\bibinfo{author}{\bibfnamefont{R.}~\bibnamefont{Banerjee}},
  \bibinfo{author}{\bibfnamefont{B.~R.} \bibnamefont{Majhi}}, \bibnamefont{and}
  \bibinfo{author}{\bibfnamefont{S.}~\bibnamefont{Samanta}},
  \bibinfo{journal}{Phys. Rev.} \textbf{\bibinfo{volume}{D77}},
  \bibinfo{pages}{124035} (\bibinfo{year}{2008}). 

\bibitem[{\citenamefont{Oh and Park}(2010)}]{OhP10}
\bibinfo{author}{\bibfnamefont{J.~J.} \bibnamefont{Oh}} \bibnamefont{and}
  \bibinfo{author}{\bibfnamefont{C.}~\bibnamefont{Park}},
  \bibinfo{journal}{JHEP} \textbf{\bibinfo{volume}{03}}, \bibinfo{pages}{086}
  (\bibinfo{year}{2010}). 

\bibinfo{author}{\bibfnamefont{P.}~\bibnamefont{Nicolini}},
  \bibinfo{author}{\bibfnamefont{A.}~\bibnamefont{Orlandi}}, \bibnamefont{and}
  \bibinfo{author}{\bibfnamefont{E.}~\bibnamefont{Spallucci}}
  (\bibinfo{year}{2011}), \eprint{arXiv:1110.5332}.

\bibitem[{\citenamefont{Garattini and Lobo}(2009)}]{Gal09}
\bibinfo{author}{\bibfnamefont{R.}~\bibnamefont{Garattini}} \bibnamefont{and}
  \bibinfo{author}{\bibfnamefont{F.~S.~N.} \bibnamefont{Lobo}},
  \bibinfo{journal}{Phys. Lett.} \textbf{\bibinfo{volume}{B671}},
  \bibinfo{pages}{146} (\bibinfo{year}{2009}). 

\bibitem[{\citenamefont{Abramowitz and Stegun~(eds.)}(1965)}]{AbS65}
\bibinfo{author}{\bibfnamefont{M.}~\bibnamefont{Abramowitz}} \bibnamefont{and}
  \bibinfo{author}{\bibfnamefont{I.~A.} \bibnamefont{Stegun~(eds.)}},
  \emph{\bibinfo{title}{Handbook of mathematical functions with formulas,
  graphs, and mathematical tables}} (\bibinfo{publisher}{Dover},
  \bibinfo{address}{New York}, \bibinfo{year}{1965}), p. \bibinfo{pages}{948}.

\bibitem[{\citenamefont{Hoyle et~al.}(2004)}]{Hoy04}
\bibinfo{author}{\bibfnamefont{C.~D.} \bibnamefont{Hoyle}}
  \bibnamefont{et~al.}, \bibinfo{journal}{Phys. Rev.}
  \textbf{\bibinfo{volume}{D70}}, \bibinfo{pages}{042004}
  (\bibinfo{year}{2004}). 

\bibitem[{\citenamefont{Nicolini}(2010)}]{Nic10}
\bibinfo{author}{\bibfnamefont{P.}~\bibnamefont{Nicolini}},
  \bibinfo{journal}{Phys. Rev.} \textbf{\bibinfo{volume}{D82}},
  \bibinfo{pages}{044030} (\bibinfo{year}{2010}). 

\bibitem[{\citenamefont{Balbinot and Barletta}(1988)}]{BaB88}
\bibinfo{author}{\bibfnamefont{R.}~\bibnamefont{Balbinot}} \bibnamefont{and}
  \bibinfo{author}{\bibfnamefont{A.}~\bibnamefont{Barletta}},
  \bibinfo{journal}{Class. Quant. Grav.} \textbf{\bibinfo{volume}{5}},
  \bibinfo{pages}{L11} (\bibinfo{year}{1988}).

\bibitem[{\citenamefont{Sprenger et~al.}(2012)\citenamefont{Sprenger, Nicolini,
  and Bleicher}}]{SNB12}
\bibinfo{author}{\bibfnamefont{M.}~\bibnamefont{Sprenger}},
  \bibinfo{author}{\bibfnamefont{P.}~\bibnamefont{Nicolini}}, \bibnamefont{and}
  \bibinfo{author}{\bibfnamefont{M.}~\bibnamefont{Bleicher}}
  (\bibinfo{year}{2012}), \eprint{arXiv:1202.1500}.

\bibitem[{\citenamefont{Kempf et~al.}(1995)\citenamefont{Kempf, Mangano, and
  Mann}}]{KMM95}
\bibinfo{author}{\bibfnamefont{A.}~\bibnamefont{Kempf}},
  \bibinfo{author}{\bibfnamefont{G.}~\bibnamefont{Mangano}}, \bibnamefont{and}
  \bibinfo{author}{\bibfnamefont{R.~B.} \bibnamefont{Mann}},
  \bibinfo{journal}{Phys.Rev.} \textbf{\bibinfo{volume}{D52}},
  \bibinfo{pages}{1108} (\bibinfo{year}{1995}). 

\bibitem[{\citenamefont{Adler et~al.}(2001)\citenamefont{Adler, Chen, and
  Santiago}}]{APS01}
\bibinfo{author}{\bibfnamefont{R.~J.} \bibnamefont{Adler}},
  \bibinfo{author}{\bibfnamefont{P.}~\bibnamefont{Chen}}, \bibnamefont{and}
  \bibinfo{author}{\bibfnamefont{D.~I.} \bibnamefont{Santiago}},
  \bibinfo{journal}{Gen.Rel.Grav.} \textbf{\bibinfo{volume}{33}},
  \bibinfo{pages}{2101} (\bibinfo{year}{2001}). 

\bibitem[{\citenamefont{Scardigli}(1999)}]{Sca99}
\bibinfo{author}{\bibfnamefont{F.}~\bibnamefont{Scardigli}},
  \bibinfo{journal}{Phys.Lett.} \textbf{\bibinfo{volume}{B452}},
  \bibinfo{pages}{39} (\bibinfo{year}{1999}). 

\bibinfo{author}{\bibfnamefont{G.}~\bibnamefont{Amelino-Camelia}},
  \bibinfo{author}{\bibfnamefont{M.}~\bibnamefont{Arzano}},
  \bibinfo{author}{\bibfnamefont{Y.}~\bibnamefont{Ling}}, \bibnamefont{and}
  \bibinfo{author}{\bibfnamefont{G.}~\bibnamefont{Mandanici}},
  \bibinfo{journal}{Class.Quant.Grav.} \textbf{\bibinfo{volume}{23}},
  \bibinfo{pages}{2585} (\bibinfo{year}{2006}). 

\bibitem[{\citenamefont{Carr et~al.}(2011)\citenamefont{Carr, Modesto, and
  Premont-Schwarz}}]{CMP11}
\bibinfo{author}{\bibfnamefont{B.}~\bibnamefont{Carr}},
  \bibinfo{author}{\bibfnamefont{L.}~\bibnamefont{Modesto}}, \bibnamefont{and}
  \bibinfo{author}{\bibfnamefont{I.}~\bibnamefont{Premont-Schwarz}}
  (\bibinfo{year}{2011}), \eprint{arXiv:1107.0708}.

\bibitem[{\citenamefont{Chen and Adler}(2003)}]{ChA03}
\bibinfo{author}{\bibfnamefont{P.}~\bibnamefont{Chen}} \bibnamefont{and}
  \bibinfo{author}{\bibfnamefont{R.~J.} \bibnamefont{Adler}},
  \bibinfo{journal}{Nucl.Phys.Proc.Suppl.} \textbf{\bibinfo{volume}{124}},
  \bibinfo{pages}{103} (\bibinfo{year}{2003}). 

\bibitem[{\citenamefont{Nicolini and Winstanley}(2011)}]{NiW11}
\bibinfo{author}{\bibfnamefont{P.}~\bibnamefont{Nicolini}} \bibnamefont{and}
  \bibinfo{author}{\bibfnamefont{E.}~\bibnamefont{Winstanley}},
  \bibinfo{journal}{JHEP} \textbf{\bibinfo{volume}{11}}, \bibinfo{pages}{075}
  (\bibinfo{year}{2011}). 
 
\bibitem[{\citenamefont{Bleicher et~al.}(2011)\citenamefont{Bleicher, Nicolini,
  Sprenger, and Winstanley}}]{BNS11}
\bibinfo{author}{\bibfnamefont{M.}~\bibnamefont{Bleicher}},
  \bibinfo{author}{\bibfnamefont{P.}~\bibnamefont{Nicolini}},
  \bibinfo{author}{\bibfnamefont{M.}~\bibnamefont{Sprenger}}, \bibnamefont{and}
  \bibinfo{author}{\bibfnamefont{E.}~\bibnamefont{Winstanley}},
  \bibinfo{journal}{Int.J.Mod.Phys.} \textbf{\bibinfo{volume}{E20S2}},
  \bibinfo{pages}{7} (\bibinfo{year}{2011}). 
 
\bibitem[{\citenamefont{Lottini and Torrieri}(2011)}]{LoT11}
\bibinfo{author}{\bibfnamefont{S.}~\bibnamefont{Lottini}} \bibnamefont{and}
  \bibinfo{author}{\bibfnamefont{G.}~\bibnamefont{Torrieri}},
  \bibinfo{journal}{Phys.Rev.Lett.} \textbf{\bibinfo{volume}{107}},
  \bibinfo{pages}{152301} (\bibinfo{year}{2011}). 

\bibinfo{author}{\bibfnamefont{G.}~\bibnamefont{Torrieri}},
  \bibinfo{author}{\bibfnamefont{S.}~\bibnamefont{Lottini}},
  \bibinfo{author}{\bibfnamefont{I.}~\bibnamefont{Mishustin}},
  \bibnamefont{and} \bibinfo{author}{\bibfnamefont{P.}~\bibnamefont{Nicolini}}
  (\bibinfo{year}{2011}), \eprint{arXiv:1110.6219}.

\bibitem[{\citenamefont{Modesto and Nicolini}(2010{\natexlab{b}})}]{MoN10a}
\bibinfo{author}{\bibfnamefont{L.}~\bibnamefont{Modesto}} \bibnamefont{and}
  \bibinfo{author}{\bibfnamefont{P.}~\bibnamefont{Nicolini}},
  \bibinfo{journal}{Phys. Rev.} \textbf{\bibinfo{volume}{D81}},
  \bibinfo{pages}{104040} (\bibinfo{year}{2010}{\natexlab{b}}).
 
\bibinfo{author}{\bibfnamefont{P.}~\bibnamefont{Nicolini}} \bibnamefont{and}
  \bibinfo{author}{\bibfnamefont{B.}~\bibnamefont{Niedner}},
  \bibinfo{journal}{Phys.Rev.} \textbf{\bibinfo{volume}{D83}},
  \bibinfo{pages}{024017} (\bibinfo{year}{2011}). 
 
\bibinfo{author}{\bibfnamefont{P.}~\bibnamefont{Nicolini}} \bibnamefont{and}
  \bibinfo{author}{\bibfnamefont{E.}~\bibnamefont{Spallucci}},
  \bibinfo{journal}{Phys. Lett.} \textbf{\bibinfo{volume}{B695}},
  \bibinfo{pages}{290} (\bibinfo{year}{2011}). 
 
\bibitem[{\citenamefont{Mureika et~al.}(2011)\citenamefont{Mureika, Nicolini,
  and Spallucci}}]{MNS11}
\bibinfo{author}{\bibfnamefont{J.}~\bibnamefont{Mureika}},
  \bibinfo{author}{\bibfnamefont{P.}~\bibnamefont{Nicolini}}, \bibnamefont{and}
  \bibinfo{author}{\bibfnamefont{E.}~\bibnamefont{Spallucci}}
  (\bibinfo{year}{2011}), \eprint{arXiv:1111.5830}.
 
\bibitem[{\citenamefont{Sprenger
  et~al.}(2011{\natexlab{a}})\citenamefont{Sprenger, Nicolini, and
  Bleicher}}]{SNB11a}
\bibinfo{author}{\bibfnamefont{M.}~\bibnamefont{Sprenger}},
  \bibinfo{author}{\bibfnamefont{P.}~\bibnamefont{Nicolini}}, \bibnamefont{and}
  \bibinfo{author}{\bibfnamefont{M.}~\bibnamefont{Bleicher}},
  \bibinfo{journal}{Class. Quant. Grav.} \textbf{\bibinfo{volume}{28}},
  \bibinfo{pages}{235019} (\bibinfo{year}{2011}{\natexlab{a}}).
 
\bibinfo{author}{\bibfnamefont{M.}~\bibnamefont{Sprenger}},
  \bibinfo{author}{\bibfnamefont{P.}~\bibnamefont{Nicolini}}, \bibnamefont{and}
  \bibinfo{author}{\bibfnamefont{M.}~\bibnamefont{Bleicher}},
  \bibinfo{journal}{Int.J.Mod.Phys.} \textbf{\bibinfo{volume}{E20S2}},
  \bibinfo{pages}{1} (\bibinfo{year}{2011}{\natexlab{b}}).

\end{thebibliography}
    \end{document}